# Title

Methods of multi-indication meta-analysis for health technology assessment: a simulation study

# Abstract

A growing number of oncology treatments, such as bevacizumab, are used across multiple indications. However, in health technology assessment (HTA), their clinical and cost-effectiveness are typically appraised within a single target indication. This approach excludes a broader evidence base across other indications. To address this, we explored multi-indication meta-analysis methods that share evidence across indications.

We conducted a simulation study to evaluate alternative multi-indication synthesis models. This included univariate (mixture and non-mixture) methods synthesizing overall survival (OS) data and bivariate surrogacy models jointly modelling treatment effects on progression-free survival (PFS) and OS, pooling surrogacy parameters across indications. Simulated datasets were generated using a multistate disease progression model under various scenarios, including different levels of heterogeneity within and between indications, outlier indications, and varying data on OS for the target indication. We evaluated the performance of the synthesis models applied to the simulated datasets, in terms of their ability to predict overall survival (OS) in a target indication.

The results showed univariate multi-indication methods could reduce uncertainty without increasing bias, particularly when OS data were available in the target indication. Compared with univariate methods, mixture models did not significantly improve performance and are not recommended for HTA. In scenarios where OS data in the target indication is absent and there were also outlier indications, bivariate surrogacy models showed promise in correcting bias relative to univariate models, though further research under realistic conditions is needed.

Multi-indication methods are more complex than traditional approaches but can potentially reduce uncertainty in HTA decisions.

# Authors

David Glynn, Centre for Health Economics, University of York. David.glynn@york.ac.uk

Pedro Saramago, Centre for Health Economics, University of York.

Janharpreet Singh, Biostatistics Research Group, Dept. of Population Health Sciences, University of Leicester. ORCID: 0000-0002-0272-3902

Sylwia Bujkiewicz, Biostatistics Research Group, Dept. of Population Health Sciences, University of Leicester

Sofia Dias, Centre for Reviews and Dissemination, University of York. ORCID: 0000-0002-2172-0221.

Stephen Palmer, Centre for Health Economics, University of York.

Marta Soares, Centre for Health Economics, University of York.

# 1. Background

Many oncology treatments are licensed for multiple indications. For example, bevacizumab is licensed in breast, colon or cervical cancers, amongst other cancer types (Garcia, Hurwitz et al. 2020). However, licensing and health technology assessment (HTA) decisions for treatments are typically made on an indication-by-indication basis, relying only on evidence from the specific "target" indication of interest. For multi-indication treatments, incorporating evidence across indications, where appropriate, could strengthen clinical and cost-effectiveness estimates, reduce uncertainty in licensing and HTA decisions, and expedite patient access (Nikolaidis, Woods et al. 2021).

A recent study (Singh, Anwer et al. 2023) examined the use of multi-indication evidence in HTA to estimate the effectiveness of bevacizumab on overall survival (OS) in advanced or metastatic cancers. The data consisted of 41 randomised controlled trials (RCTs) across seven cancer types, all trials reporting log hazard ratios (LHR) for progression free survival (PFS) and 36 also reporting LHRs for OS. The study applied univariate models synthesising OS effects and bivariate surrogacy synthesis models, which established indication-specific surrogate relationships between the treatment effects on PFS and OS and synthesised surrogacy parameters. Models explored alternative sharing assumptions for the syntheses across indications. Independent parameters (IP), imposing no sharing of information; common parameters (CP), imposing full sharing of information; random parameters (RP), imposing partial sharing of information through an exchangeability assumption; or mixture models, also imposing partial sharing by allowing each indication to either be independent or to share information (using CP or RP), with the probability of sharing estimated from the data.

The data analysed in Singh et al (Singh, Anwer et al. 2023) showed small but consistent treatment effects on OS and PFS across studies, which did not seem to differ across indications (Singh, Anwer et al. 2023). The application of univariate sharing models to this dataset, particularly under the common parameter assumption, significantly increased the precision of OS estimates compared to independent analysis. More complex mixture and bivariate models did not improve precision or fit and, in some cases, increased uncertainty. The limited performance of the more complex models was an unexpected finding but could potentially be explained by the limited between indication heterogeneity in the dataset.

To support the broader application of multi-indication synthesis approaches a simulation study in which estimates can be compared against known true values of the estimand is needed, to evaluate the performance of these methods. Further, it is necessary to assess the generalisability of the Singh et al case study application to the variety of data structures encountered within HTA, including contexts of greater heterogeneity. More complex synthesis methods may perform better under larger levels of heterogeneity, in the presence of potential outliers and where evidence on the LHR for OS in the target indication is sparse, absent or biased (Bujkiewicz, Achana et al. 2019).

The primary aim of this study was to assess the performance and impact of multi-indication synthesis methods compared to standard practices where evidence from other indications is excluded. Using simulation, we assessed these methods across various data structures considered relevant to HTA, focusing on their ability to predict treatment effects on OS for a specific target indication.  Additionally, our work explored the conditions under which more complex synthesis methods, such as mixture and surrogacy models, might reduce bias and improve precision relative to simpler models.

To simulate the data we used a previously developed data-generating model (DGM) (Erdmann, Beyersmann et al. 2025), based on a multi-state model (MSM) of cancer progression with three states: pre-progression, progressed disease, and death. This MSM model, well-established in oncology, has been used to represent natural history (Bullement, Cranmer et al. 2019, Woods, Sideris et al. 2020, Cheung, Albert et al. 2022), quantify treatment effects (Jansen, Incerti et al. 2023), simulate surrogate relationships (Weber and Titman 2019), and simulate trial data for sample size calculations (Erdmann, Beyersmann et al. 2025). However, the LHR for PFS and OS analysed by the synthesis models are not explicit parameters in the DGM but can be derived to obtain a joint distribution. Because the multi-indication synthesis models assume univariate or bivariate normality, the model specification differs between the DGM and the synthesis models. However, this is desirable as it results in realistic conditions for our simulation evaluation.

# 2. Multi-indication synthesis models

The following sections describe the multi-indication synthesis models considered in this paper; further details can be found in Singh et al (2023).

## 2.1. Univariate non-mixture models

In univariate models, the observed log hazard ratios (LHR) for overall survival (OS) in study i and indication j, $Y_{OS,ij}$, are assumed to be normally distributed around their true value with standard errors $\sigma_{OS,ij}$.

$$Y_{OS,ij} \sim N(d_{OS,ij}, \sigma^2_{OS,ij}).$$

Random effects describe study results within each indication, meaning that the study level effects were assumed normally distributed, with mean, $D_{OS,j}$, representing the pooled effect for indication j. $\tau_{OS,j}$ was the between-study, within-indication, standard deviation:

$$d_{OS,ij} \sim N(D_{OS,j}, \tau^2_{OS,j}).$$

Assuming independent standard deviation parameters across indications (as in (Singh, Anwer et al. 2023)) is likely to require substantial data within each indication for precise estimation. Hence, we also explore the alternative assumption that this parameter is common across indications ($\tau_{OS,j} = \tau_{OS}$), assigning a weakly informative half normal prior to the common parameter $|N(0, 0.5^2)|$ (Röver, Wandel et al. 2019).

To encode how information on $D_{OS,j}$ is related across indications (determining the sharing of information across indications), three different assumptions were explored (using either independent or common standard deviation parameters). Independent parameter (IP) model, imposing no sharing between indications. Common parameter (CP) model, assuming equality across indications, i.e. $D_{OS,j} = D_{OS}$, in this way imposing maximal sharing of information. Random parameter (RP) model, imposing partial sharing by assuming $D_{OS,j}$ to be exchangeable across indications, $D_{OS,j} \sim N(m_d, \varepsilon_d^2)$. Six univariate models were therefore examined: $IP_\tau$, $IP_{\tau j}$, $CP_\tau$, $CP_{\tau j}$, $RP_\tau$ and $RP_{\tau j}$. Table 1 summarises how LHR OS in the target indication was predicted from each model. For all models (except CP), this depends on availability of LHR OS data for the target indication. IP models rely on indication-specific evidence, and without this, predicted estimates of the LHR OS cannot be obtained. Because CP models use the common effect, in both cases the LHR of OS can be predicted with or without LHR OS data in the target indication. In the RP model, shrunken indication-specific estimates were used to predict OS when OS data was present and the predictive distribution, $D_{OS,pred} = N(m_d, \varepsilon_d^2)$ when OS data was absent.

## 2.2. Univariate mixture models

Mixture models consider the effects in each indication to be either independent or from a shared distribution. The mixture probability reflected the probability that the effect in indication j came from the shared distribution, based on similarity with the other indication-level effects (i.e., a large probability would reflect strong similarity between effects). This was controlled by an indicator Bernoulli variable $c_j$, which assumed the value of 1 for shared and 0 for independent. The Bernoulli probability parameter was estimated from the data. If the sharing component assumed common parameters across indications, this resulted in the mixture common and independent parameter (MCIP) model:

$$D_{OS,j} = \begin{cases} D_{OS}, & c_j = 1 \\ \sim N(0, 10^2), & c_j = 0 \end{cases}.$$

If the sharing component used random parameters across indications (i.e. exchangeable), this resulted in the mixture random and independent parameter (MRIP) model:

$$D_{OS,j} = \begin{cases} \sim N(m_d, \varepsilon_d^2), & c_j = 1 \\ \sim N(0, 10^2), & c_j = 0 \end{cases}.$$

As with non-mixture models, the within-indication heterogeneity parameter was here also assumed either independent or common. This resulted in four univariate mixture model estimates: $MCIP_\tau$, $MCIP_{\tau j}$, $MRIP_\tau$ and $MRIP_{\tau j}$. The predicted effect in the target indication from each model was generated from the sharing component of the mixture model, the common CP for MCIP and RP for MRIP (see Table 1).

## 2.3. Bivariate (surrogacy) models

The bivariate models applied in (Singh, Anwer et al. 2023) and further examined in this work extend Daniels and Hughes to consider multiple indications (Daniels and Hughes 1997). In these models, a within-trial component describes the relationship between the treatment effects on a surrogate endpoint (PFS) and a final clinical outcome (OS) within an individual study, i:

$$\begin{pmatrix} Y_{PFS,ij} \\ Y_{OS,ij} \end{pmatrix} \sim N\left( \begin{pmatrix} d_{PFS,ij} \\ d_{OS,ij} \end{pmatrix}, \begin{pmatrix} \sigma^2_{PFS,ij} & \sigma_{PFS,ij}\sigma_{OS,ij}\rho_w \\ \sigma_{PFS,ij}\sigma_{OS,ij}\rho_w & \sigma^2_{OS,ij} \end{pmatrix} \right) \quad (1)$$

where $Y_{PFS,ij}$ $Y_{OS,ij}$ represent the observed LHR for PFS and OS, respectively, with standard errors $\sigma_{PFS,ij}$ and $\sigma_{OS,ij}$, and within study correlation, $\rho_w$. Correlation was assumed common across studies and indications and assigned a weakly informative prior $\rho_w \sim U(0,1)$. A between-trial component describes the relationship between LHR OS and LHR PFS between studies (in the same indication, in this case), where a linear surrogate relationship is assumed between the true treatment effects on the surrogate endpoint $d_{PFS,ij}$ and the final outcome $d_{OS,ij}$:

$$d_{OS,ij} \sim N(\gamma_{0j} + \gamma_{1j} d_{PFS,ij}, \psi_j^2), \quad (2)$$

where $\gamma_{0j}$ and $\gamma_{1j}$ are the intercept and slope respectively in indication j. $\psi_j^2$ is the conditional variance of the relationship which captures the extent to which a trial-specific LHR OS can be predicted from LHR PFS.

We examined sharing in all parameters in the surrogate relationship ($\gamma_{0j}$ and $\gamma_{1j}$ and $\psi_j^2$). Two alternative sharing relationships were considered. A bivariate CP model (Bi-CP) assumed each surrogacy parameter as common across indications, and a bivariate RP model (BI-RP) assumed each as exchangeable, i.e. $\gamma_{0j} \sim N(\beta_0, \xi_0^2)$, $\gamma_{1j} \sim N(\beta_1, \xi_1^2)$ and $\psi_j \sim |N(0,h)|$. Hyperparameters were assigned vague prior distributions.

Prediction of LHR OS for the target indication is summarised in Table 1 and requires estimates of the surrogate relationship from equation (2) and an estimate of LHR PFS for the target indication, $D_{PFS,j*}$. The latter comes from the univariate models described in sections 2.1 and 2.2 and can itself use different sharing assumptions. Following Singh et al, we considered "unmatched" and "matched" estimates. Matched estimates used the same sharing assumption (either CP or RP) for both the surrogate relationship and the univariate LHR PFS model. Unmatched estimates use a univariate IP model for LHR PFS data, and a sharing model for the surrogate parameters. For the LHR PFS estimate, we explored common or independent τ for the random effects within indications. This resulted in eight estimates from bivariate models being examined here considering Bi-CP or Bi-RP, matched or unmatched and with either independent or common within-indication heterogeneity parameter.

*Table 1: Synthesis models investigated and prediction of LHR on OS in the target indication from each model. Each listed model was ran twice, assuming common or independent between-study heterogeneity parameter. Note that the prediction of the expected indication-specific effects from bivariate models do not use the conditional variance as this parameter captures variation across trials which is deemed unwarranted. j\* = target indication; IP = independent parameters; CP = common parameters; RP = random parameters; MCIP =*

*mixed common and independent parameters; MRIP = mixed random and independent parameters; Bi = bivariate; PFS = progression free survival; OS = overall survival.*

| Model summary focussing on between-indication sharing assumption | | Prediction of LHR OS in target j* | |
|---|---|---|---|
| ID | Description | OS in target indication | No OS in target indication |
| IP | LHR OS: Independent | $D_{OS,j*}$ | - |
| CP | LHR OS: Common | $D_{OS}$ | |
| RP | LHR OS: Exchangeable | $D_{OS,j*}$ | $D_{OS,pred}$ |
| MCIP | LHR OS: Mixture of common and independent | $D_{OS}$ | |
| MRIP | LHR OS: Mixture of exchangeable and independent | $D_{OS,j*}$ | $D_{OS,pred}$ |
| Bi-CP Unmatched | LHR PFS: independent | $\gamma_0 + \gamma_1 D_{PFS,j*}$ | |
| | Surrogacy parameters: common | | |
| Bi-RP unmatched | LHR PFS: independent | $\gamma_{0j*} + \gamma_{1j*} D_{PFS,j*}$ | $\gamma_{0pred} + \gamma_{1pred} D_{PFS,j*}$ |
| | Surrogacy parameters: exchangeable | | |
| Bi-CP matched | LHR PFS: common | $\gamma_0 + \gamma_1 D_{PFS}$ | |
| | Surrogacy parameters: common | | |
| Bi-RP matched | LHR PFS: exchangeable | $\gamma_{0j*} + \gamma_{1j*} D_{PFS,j*}$ | $\gamma_{0pred} + \gamma_{1pred} D_{PFS,j*}$ |
| | Surrogacy parameters: exchangeable | | |

# 3. Simulation study methods

The simulation study was designed and reported using the "ADEMP" approach -- Aims, Data-generation, Estimands, Methods, and Performance measures (Morris, White et al. 2019).

## 3.1.  Aims

- Investigate if univariate (non-mixture) multi-indication sharing methods improve inferences over non-sharing methods by increasing precision, calibrating uncertainty and maintaining low bias.

- Identify when different univariate sharing assumptions (common or exchangeable parameters) are appropriate and lead to significant precision gains, low bias and well calibrated descriptions of uncertainty.

- Explore when mixture models may reduce bias and calibrate uncertainty compared to non-mixture models.

- Explore when bivariate models may reduce bias and calibrate uncertainty over univariate models. This is exploratory and will be investigated under ideal conditions for surrogacy.

## 3.2. Data generation

Generating multi-indication datasets requires definition of the following elements: the DGM, the sets of parameter values examined, features of datasets (such as number of indications and study sample size), and the simulation process. We describe each of these elements in turn.

### Data generating mechanism

The DGM uses the MSM from Erdmann et al shown in Figure 1 (Erdmann, Beyersmann et al. 2025). This is a 3-state model that determines a structural relationship between progression and mortality. Patients start in the pre-progression state (state 0), where they are at risk of progression and death. The hazard of progression is $\lambda_{01}M_j$, with $\lambda_{01}$ representing the hazard under current practice and the multiplier, $M_j$, representing how treatment impacts on $\lambda_{01}$ in indication j (slowing progression if $M_j < 1$, accelerating it if $M_j > 1$). The pre-progression hazard of death is $\lambda_{02}$. The multiplier $\Delta > 1$ reflects the increase in mortality hazard after progression. Exponential hazards were assumed for all transitions. PFS is defined as time in state 0, and OS as the time spent in states 0 and 1. The MSM is used to (jointly) simulate time to progression and death for individual patients in a clinical study (Erdmann, Beyersmann et al. 2025), which are subsequently used to define PFS and OS.

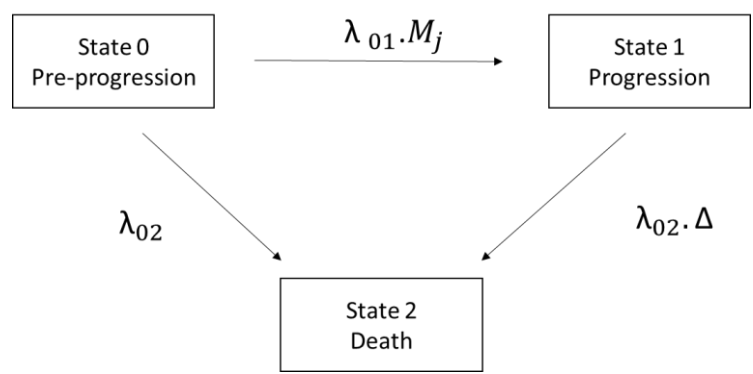

Figure 1: Data generating mechanism: three state MSM defining the relationship between PFS and OS in indication j. $\lambda_{01}$ indicates the rate of progression, $\lambda_{02}$ is the rate of pre-progression death, $\Delta$ is the increase in mortality that results from progression. $M_j$ is an indication specific multiplier which encodes how the new treatment impacts on the rate of progression. MSM = multi-state model; PFS = progression free survival; OS = overall survival.

### Parameter value sets

The parameter value sets used in the MSM are summarised in Table 2; we here describe how those values were obtained. Estimates for MSM parameters unrelated to treatment, i.e. $\lambda_{01}$ and $\Delta$, were derived from data on the control arms of the largest RCTs for bevacizumab (Singh, Anwer et al. 2023), and supported by an estimate of $\lambda_{02}$ from Jansen et al (Jansen, Incerti et al. 2023). The control arms in most of these studies consisted of treatment with chemotherapy representing the absence of targeted treatment. Between and within-indication heterogeneity on these parameters were not considered, so as to isolate the effects of treatment effect heterogeneity. Values were therefore derived independently by study, and averaged to retrieve an overall estimate. Further details can be found in Appendix A1.

Treatment effects, M, were assumed to exhibit heterogeneity, both between- and within-indications. The treatment effects were described using distributions and simulated values were drawn from the nested log Normal distributions shown in the equations below:

$$\ln(M_j) \sim N(\ln(\mu_M), \sigma_b),$$

$$\ln(M_{ji}) \sim N(\ln(M_j), \sigma_w),$$

where $M_j$ is the treatment effect for indication j, sampled from a log Normal distribution with mean $\mu_M$ (the central treatment effect estimate across all indications) and standard deviation $\sigma_b$ representing between-indication heterogeneity. $M_{ji}$ is the treatment effect in study i (within indication j), sampled from a log Normal distribution using the sampled $M_j$ as its mean value and a pre-defined value of $\sigma_w$ representing the within-indication standard deviation.

A value of 0.6 was used for $\mu_M$ to represent a moderate treatment effect in oncology HTA, the heterogeneity parameters, $\sigma_b$ and $\sigma_w$, were defined using coefficients of variation (CV = $\sigma$/abs($\mu_M$)) to ensure that conclusions are as independent as possible from the specific value of $\mu_M$. CV values examined were 0%, 7%, 15%, 30% and 50%. A CV of 50% (the maximum considered) corresponds to a distribution where approximately 2.5% of simulated $M_j$ values exceed 1, indicating harm. Since this study focusses on approved indications under regulatory standards, higher probabilities of a treatment being harmful in a particular indication are unlikely.

The way treatment effect heterogeneity was defined means that treatment effects vary across indications but are centred around a common value, $\ln(\mu_M)$. Exchangeability can therefore be considered plausible. To test alternative assumptions, we ran scenario analyses introducing an outlier indication with a 'divergent' treatment effect value, not sampled from the same process. The treatment effect in the outlier indication was defined as being $\mu_M = 1.96\sigma_b$, described as a 'moderate' outlier, with M drawn from the upper 95% interval for between indication heterogeneity. Additionally, an 'extreme' outlier was defined as $\mu_M = 6\sigma_b$, with M drawn from the upper 99.99% interval. We explored the outlier being a non-target indication and, to evaluate the advantages of bivariate methods, we also explored a scenario where the target indication is the outlier.

### Features of the multi-indication datasets

The number of indications and RCTs per indication in each scenario was based on the features of the bevacizumab dataset (Singh, Anwer et al. 2023). Three cases were defined: a "large" evidence base with 8 indications representing the most developed dataset for bevacizumab, a "medium" base with 6 indications, and a "small" base with 4 indications. To reflect the HTA context, one indication was defined as the target, and was assumed to include only one study which reported either PFS only or both PFS and OS.

The follow-up duration for all studies was set so that 80% of OS events occurred in the control arm. Following the studies in the Singh et al case study, sample sizes were chosen so that each study had the power to detect a 0.7 hazard ratio with a 5% Type-1 error rate and 90% power (10% Type-2 error rate). The power calculation formula used was Lachin-Foulkes as implemented in the 'gsDesign' package in R (Anderson 2024).We assumed an exponential rate of OS events and a two-arm balanced design, zero dropout and instant accrual.

|  | **Description** |
|---|---|
| **MSM parameter mean values** | $\mu_{\lambda 01} = 0.097$, $\mu_{\lambda 02} = 0.01$, $\mu_\Delta = 6.32$, $\mu_M = 0.6$ on natural scale |
| **Scenarios** | **Description** |
| **Variance parameter describing heterogeneity in the treatment effect, M** | In all scenarios, between- and within-indication heterogeneity in M was varied assuming coefficient of variation (CV) values of 0%, 7%, 15%, 30%, 50%. Other parameters ($\mu_{\lambda 01}$, $\mu_{\lambda 02}$, $\mu_\Delta$) were assumed independent by indication. These were not assumed heterogeneous within-indications. Values were therefore kept constant across studies in the same indication. |
| **Outlier indications** | **No outlier indications**: All indications, including the target, were assumed exchangeable. $M_j$ values were randomly sampled from a common distribution based on a pre-specified common mean and the between-indication heterogeneity parameter value.<br><br>**One moderate (non-target) outlier indication**: Same as above for all indications except the second largest indication, for which the M value was fixed at the 95% percentile ($1.96\sigma_b$) of the between indication heterogeneity distribution, rather than being randomly sampled.<br><br>**One extreme (non-target) outlier indication**: Same as above but the outlier indication had an M that was $6\sigma_b$ away from the mean.<br><br>**Outlier target indication**: Same as the no outlier indication scenario, but the M for the target indication was centred at the 95% percentile ($1.96\sigma_b$) of the distribution describing between indication heterogeneity in M. |
| **Size of evidence base** | **Small:** 4 indications. 3 indications reporting PFS and OS for 3, 2 and 1 study respectively. 1 indication reporting PFS only.<br><br>**Medium:** 6 indications. 5 indications reporting PFS and OS for 7, 3, 3, 2 and 1 study respectively. 1 indication reporting PFS only.<br><br>**Large**: 8 indications. 7 indications reporting PFS and OS for 9, 8, 6, 3, 2, 1 and 1 study respectively. 1 indication reporting PFS only. |
| **Target indication** | **With OS:** target indication had both OS and PFS data.<br><br>**Without OS** [Results only presented in appendix]**:** target indication had only PFS data. |

*Table 2: Design factors for the simulation study. The full set of design factors factorially varied in the simulation study resulted in (5 x 5 x 4 x 3 x 2 =) 600 simulation scenarios. OS = overall survival; PFS = progression free survival; CV = coefficient of variation for parameter M; MSM = multi state model, $\sigma_b$ = standard deviation of between indication heterogeneity.*

### Simulation process

As described above, study-level parameter values on M were sampled from their distributions. This generates a set of values with which to run the DGM simulation, generating simulated outcomes for individual patients (i.e. considering sampling uncertainty) according to a hypothetical trial with a set of defined features. The simulation used the 'simIDM' package in R (Rufibach 2023), which implements the MSM model, generating individual values of time to progression and death. This was based on a nested set of competing risks experiments in a continuous time framework that implements the Erdmann et al model (Erdmann, Beyersmann et al. 2025). The simulated PFS and OS individual patient data from each study were analysed by fitting Cox proportional hazard survival models to each outcome separately to estimate the LHRs for both outcomes. The Cox model was used because this is ubiquitous in practice. However, as shown by Erdmann et al (Erdmann, Beyersmann et al. 2025), LHR OS exhibits non-proportional hazards meaning it varies over time (except under very unrealistic assumptions). Therefore, the proportional hazards assumption can only hold approximately for OS.

## 3.3. Estimand

An estimand is the specific quantity a study seeks to infer (Lundberg, Johnson et al. 2021, Kahan, Hindley et al. 2024). In oncology HTA, a key estimand is the treatment's relative effect on OS in the target indication i.e. the LHR OS. As noted above, LHR OS typically varies over time, so following Berry et al (1991), we defined the estimand as the expected LHR OS over study duration in the target indication. This was similar to restricted expectation and was based on the MSM parameters, $\lambda_{01}$, $\lambda_{02}$, $\Delta$ and M. The true estimand value was calculated as a weighted average of the true LHRs at discrete time points (1.5 day intervals) weighted by the proportion of events expected within each intervals (Berry, Kitchin et al. 1991, Mukhopadhyay, Huang et al. 2020).

## 3.4. Methods

We simulate 1000 multi-indication datasets as described in section 3.2. Each of the statistical models described in section 2 were fit to these datasets, using Markov Chain Monte Carlo (MCMC) in R JAGS. Three chains, a 50,000 burn in and 150,000 iterations were considered (Plummer, Stukalov et al. 2023). Two criteria were used to assess MCMC convergence: Option 1 required $\hat{R}$ < 1.1 for all parameters (Gelman, Carlin et al. 1995), while Option 2 required $\hat{R}$ < 1.1 for only the model components used for prediction (see appendix A3). Runs that didn't meet Option 2 were dropped. All analysis was carried out using the Viking Cluster, a high-performance computer resource provided by the University of York. The simulation code is available at https://github.com/david-glynn/Multi-indication-simulation.

## 3.5. Performance measures

The performance of each multi-indication analyses model was evaluated using the following metrics (on the log hazard scale): bias (average error), coverage (the proportion of times that the true effect lies within the 95% credible interval which should be 0.95) and empirical standard error (variance of the estimate across iterations) (Morris, White et al. 2019). To keep the paper concise, the latter metric is only shown in appendix. The strength of sharing for each method was assessed using the "splitting SE ratio", which is the ratio of the SE estimate from a sharing model to that from the non-sharing model (IP$_\tau$) (Nikolaidis 2020). A value less than 1 indicates a gain of strength. These metrics were chosen to understand the statistical reliability the methods and the magnitude of potential reductions in uncertainty. Monte Carlo errors were calculated for each metric (Morris, White et al. 2019).

# 4. Results

We compared the performance of 18 synthesis models (each of the approaches in Table 1 evaluated twice assuming common or independent random-effect heterogeneity parameters) across 600 scenarios, defined by factorially varying the design factors in Table 2. To keep the results section succinct the figures shown in main text are selected, for relevance with the full results available in Appendix A4.

## 4.1. Univariate (non-mixture) models

In this subsection, we examine simulation results for the six univariate non-mixture models considered -- the IP, CP and RP models, assuming either $\tau_j$ or $\tau$ -- applied to the case where there is OS data in the target indication. The models showed a high convergence rate (>95%) across all analyses (for both convergence options 1 and 2). Figure 2 displays the performance metrics (y-axis) for the large and small evidence-bases in panels A and B, respectively. The left side of each panel shows scenarios without outlier indications, while the right side shows scenarios with one extreme (non-target) outlier indication. The x-axis is divided into five sections, corresponding to levels of between-indication heterogeneity in M, ranging from CV = 0% to 50% (top x-axis). Within each of these sections, scenarios vary by within-indication heterogeneity, from CV = 0% to 50% (bottom x-axis).

Panel A: Scenario with a large evidence-base.

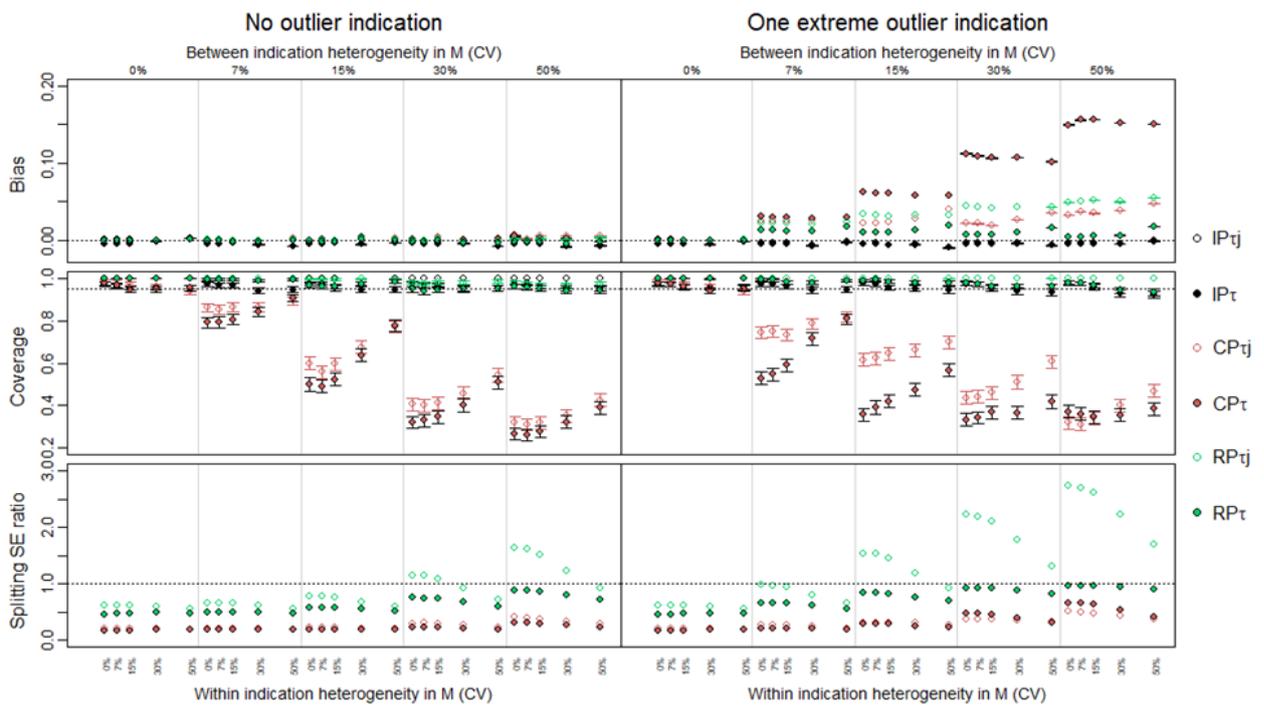

Panel B: Scenario with a small evidence-base.

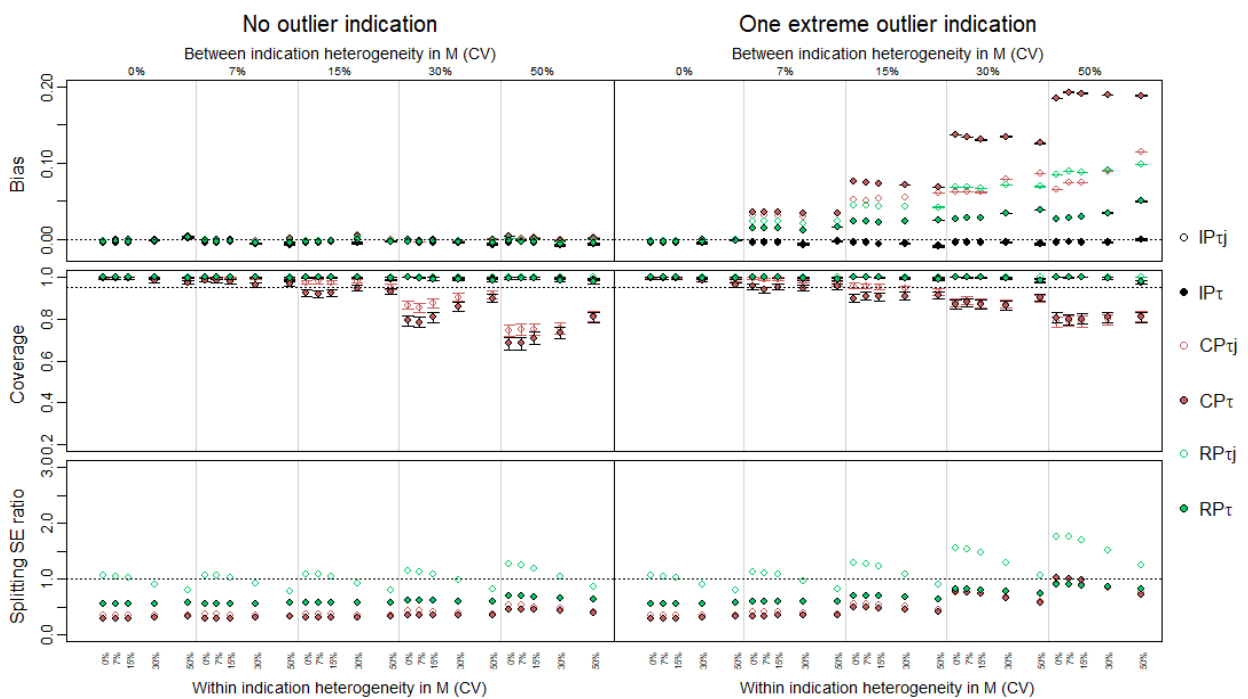

Figure 2: Performance measures comparing univariate non-mixture models with OS in the target indication. Results are shown for technologies with large (Panel A) and small (Panel B) evidence bases. The left panels show performance when there are no outlier indications, the right panels show performance for when the second largest indication is an outlier. Results are shown for all combinations of within indication heterogeneity (CV 0%, 7%, 15%, 30%, 50%) and between indications (CV 0%, 7%, 15%, 30%, 50%). CV = coefficient of variation; IP = independent parameters; CP = common parameters; RP = random parameters; τ = common within indication heterogeneity; τj = independent within indication heterogeneity.

We first examine the case with no outlier indications (left plots in panels A and B, Figure 1). Results show that all univariate models examined were approximately unbiased (bias values close to zero), across all levels of within- and between-indication heterogeneity. This was expected as, in the absence of outliers, the DGM implies exchangeability in treatment effects across all indications, including the target. This means that, even under heterogeneity, the indication estimates are centred around a common value. This is well captured by the multi-indication models.

In the large evidence-base (panel A), CP models were "well-calibrated" for coverage at low levels of between-indication heterogeneity (CV≤7%) i.e. the 95% interval included the true value ≈ 95% of the time. However, they were "overconfident" at higher heterogeneity levels i.e. they included the true value more than 95% of the time. RP models were anticipated to adjust for heterogeneity and remain well calibrated across the range of heterogeneity levels. However, they were underconfident at low heterogeneity, and improved only at higher levels of heterogeneity. This may be due to identifiability issues for the within- and between-indication heterogeneity parameters, likely caused by the fact that data are sparse at the indication-level leading to undue dominance of the "weakly informative" prior.

Sharing models showed lower splitting SE than non-sharing (IP) models in the large evidence-base (panel A) meaning they borrowed more strength. CP models shared more strongly than RP models, with RP models showing only modest precision gains at higher levels of heterogeneity (CV≥30%). In the small evidence base (panel B), the splitting SE increased for all models, and the overconfidence in coverage seen in CP models at high heterogeneity was less pronounced.

We now examine the case where there is an outlier (non-target) indication (right plots in panels A and B, Figure 1). As expected, the IP model remained unbiased but the sharing models showed bias. By virtue of the way the outlier was defined, the magnitude of bias depended on the level of between-indication heterogeneity. Among sharing models, $RP_\tau$ showed the least bias because its shrunken estimate gives more weight to the indication-specific data, especially in cases of conflict. The effect of the assumption of common vs. independent heterogeneity parameters on bias differed between RP and CP models: common heterogeneity imposed higher bias with CP and lower bias in RP. The effect of such an assumption is complex but seems to increase the precision of smaller indications, and in CP this seems to increase the influence of the outlier in overall estimates, while in RP this seems to decrease the shrinkage of the target indication and in this way reduce bias.

The coverage and splitting SE results showed that, with an outlier, the strength of sharing of analyses models is reduced especially with RP models, which inflate the between-indication heterogeneity to account for the divergence in the result of the outlier indication.

Overall, results for data structures without OS data in the target indication (Appendix A4.1.2) were similar to those presented here, except for RP models. In the absence of OS data, RP models use the predictive distribution to generate an estimate for the indication, instead of the shrunken estimate used when there is OS data (Table 1). This leads to higher uncertainty and introduces potential for bias when there is an outlier amongst the non-target indications.

## 4.2. Exploratory analyses of univariate mixture models under ideal conditions

According to convergence options 1 and 2, mixture models converged well (<5% non-convergence) except for the MCIP models when used to analyse small evidence bases (up to 32% non-convergence). The lack of convergence was associated with cases where the posterior sample space entered the region where $c_j = 0$ for all j. In this region, the common effect cannot be estimated, and its value is therefore drawn from its prior. This generates discontinuity in the posterior chain and results in a large value for $\hat{R}$.

To explore when the mixture models might overcome the limitations of non-mixture models, we restrict the presentation of results in the main paper to models that assume common within-indication heterogeneity parameter, we include only datasets with an outlier indication, and we examine model performance only in terms of bias. Full results are, however, available in appendix A4.1. Figure 3 compares bias for MCIP and MRIP with their non-mixture counterparts CP and RP, for datasets with a moderate (left side panel) and an extreme outlier (right side panel). Across all scenarios examined, mixture models presented no bias improvement over the RP model, although RP already presents relatively low levels of bias. MRIP presents similar results to its non-mixture counterpart, RP, across all scenarios. In contrast, the MCIP could only reduce bias in relation to CP when the outlier was extreme, when the evidence base was medium or large, and under high between-indication heterogeneity scenarios (≥30% CV). This pattern also held when there was no OS in the target indication.

Panel A: Technology with large evidence base

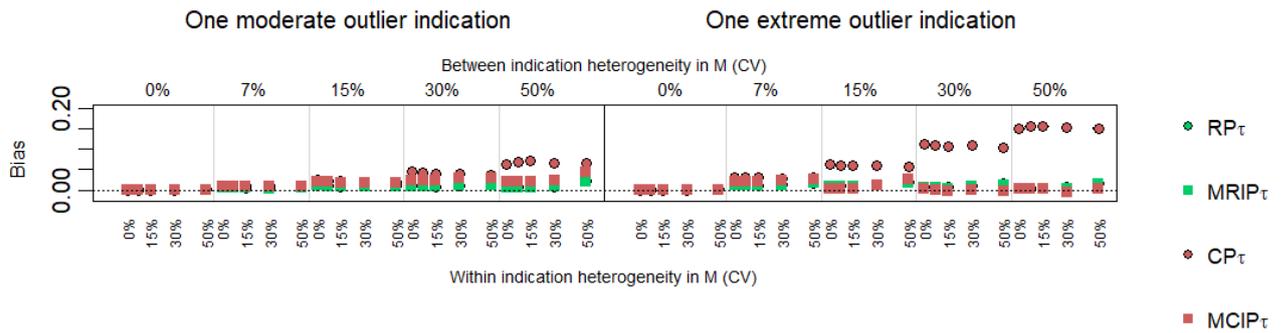

Panel B: Technology with medium sized evidence base

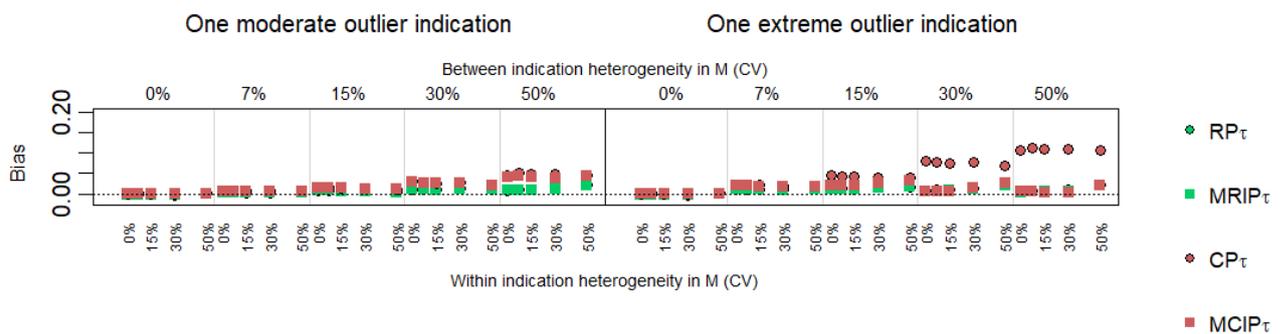

Panel C: Technology with small evidence base

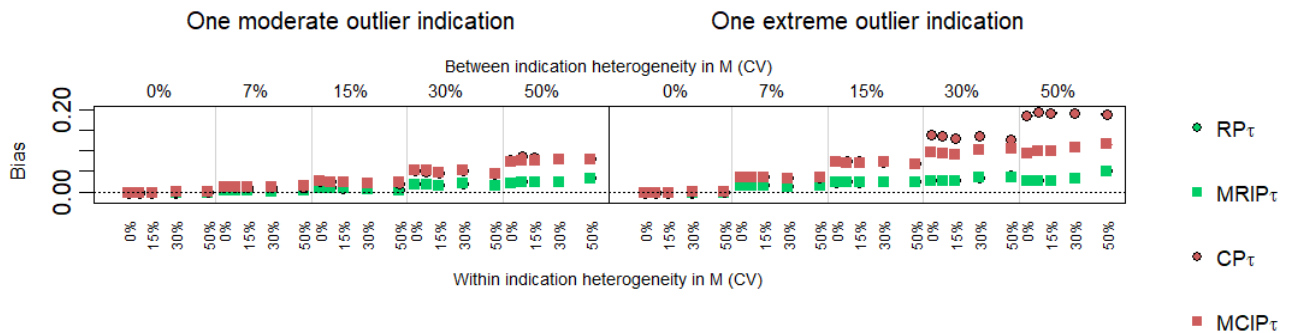

*Figure 3: Comparing bias for all univariate mixture models and $RP_\tau$ with OS in the target indication. Results are shown for a large (Panel A), medium (Panel B) and small (Panel C) evidence base. The left panels show performance when the second largest indication is a moderate outlier, the right panels show results with an extreme outlier indication. Results are shown for all combinations of within indication heterogeneity (CV 0%, 7%, 15%, 30%, 50%) and between indications (CV 0%, 7%, 15%, 30%, 50%). CV = coefficient of variation; CP = common parameters; MCIP = mixed common and independent parameters; RP = random parameters; MRIP = mixed random and independent parameters; τ = common within indication heterogeneity.*

## 4.3. Exploratory analysis of bivariate (surrogacy) models under ideal conditions

All surrogacy models converged well under the criteria in option 2. However, RP surrogacy models showed severe converge issues under the criteria in option 1 (in more than 99% of runs), indicating difficulties with the identification of parameters not used for prediction which may limit the reliability of these models.

For clarity we only present results here for unmatched Bi-CP and Bi-RP compared to the univariate models IP and RP, all under the assumption of common heterogeneity parameter, and applied to the case where there is OS in the target indication (full results are available in Appendix A4.1). The matched bivariate models are presented in the appendix only because in cases where bias is small the strength of sharing is weak (Bi-RP), or the coverage is inappropriate (Bi-CP). Further, the use of independent within-indication heterogeneity parameters was omitted at it produces highly uncertain LHR PFS predictions leading to highly uncertain OS predictions.

Figure 4 presents results with a large and small evidence-base, and with a non-target indication as outlier (extreme outlier) and with the target indication as outlier (moderate outlier). In the case where the non-target indication is an outlier (left-side panel), surrogacy models generate OS predictions with low level bias, suggesting that the IP estimate of LHR PFS is unbiased and that the surrogate relationship exists and is identified reliably. Bi-CP borrowed more strongly than Bi-RP. Critically, both models, however, showed little improvement in precision (splitting SE) compared to not-sharing, IP, except for the Bi-CP model under high between-indication heterogeneity. This limits the potential value of the sharing on surrogate relationships.

Where the target indication is the outlier (right-side panel), RP becomes biased even when there is OS in the target indication but both Bi-RP and Bi-CP were able to resolve such bias. When there is OS in the target indication, neither bivariate model reduced uncertainty relative to IP except for the Bi-CP when applied to a large evidence base and under lower within-indication heterogeneity. This, again, limits the value of these models when OS is available. In the absence of OS data on the target indication (see appendices A4.1.2 and A4.2.2), IP estimates are not available and bivariate models are the only option which provides an unbiased estimate for LHR OS.

Panel A: Technology with large evidence base

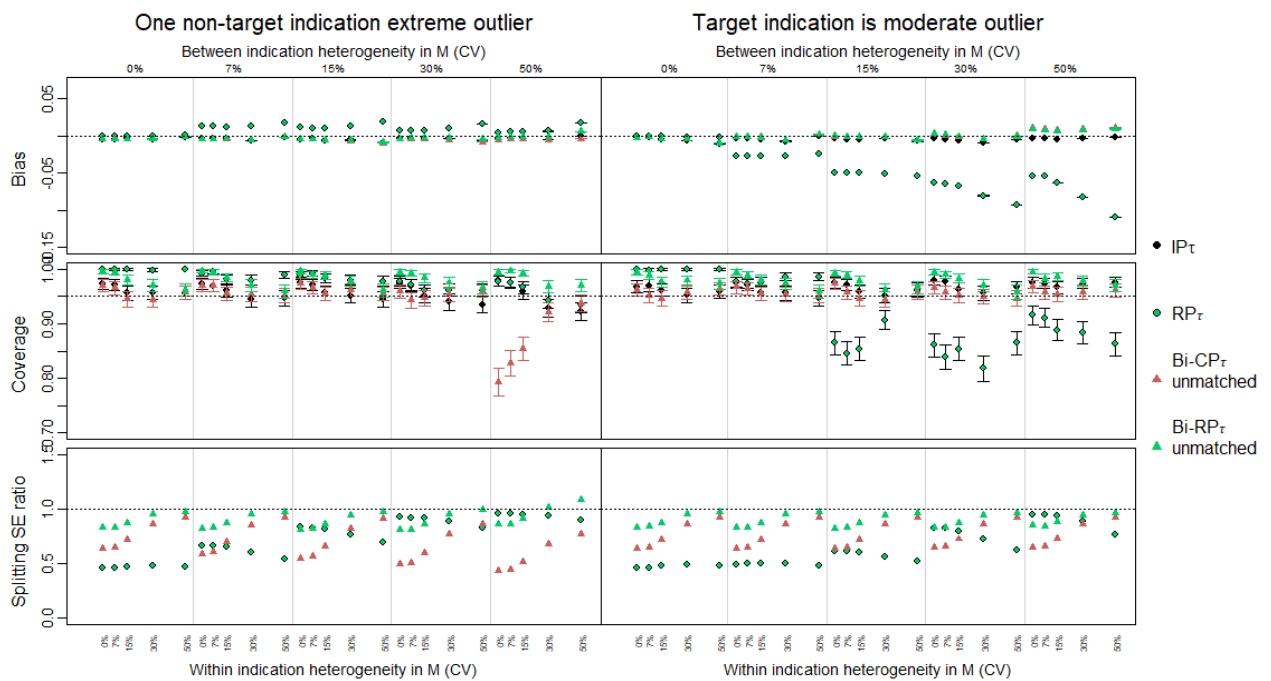

Panel B: Technology with small evidence base

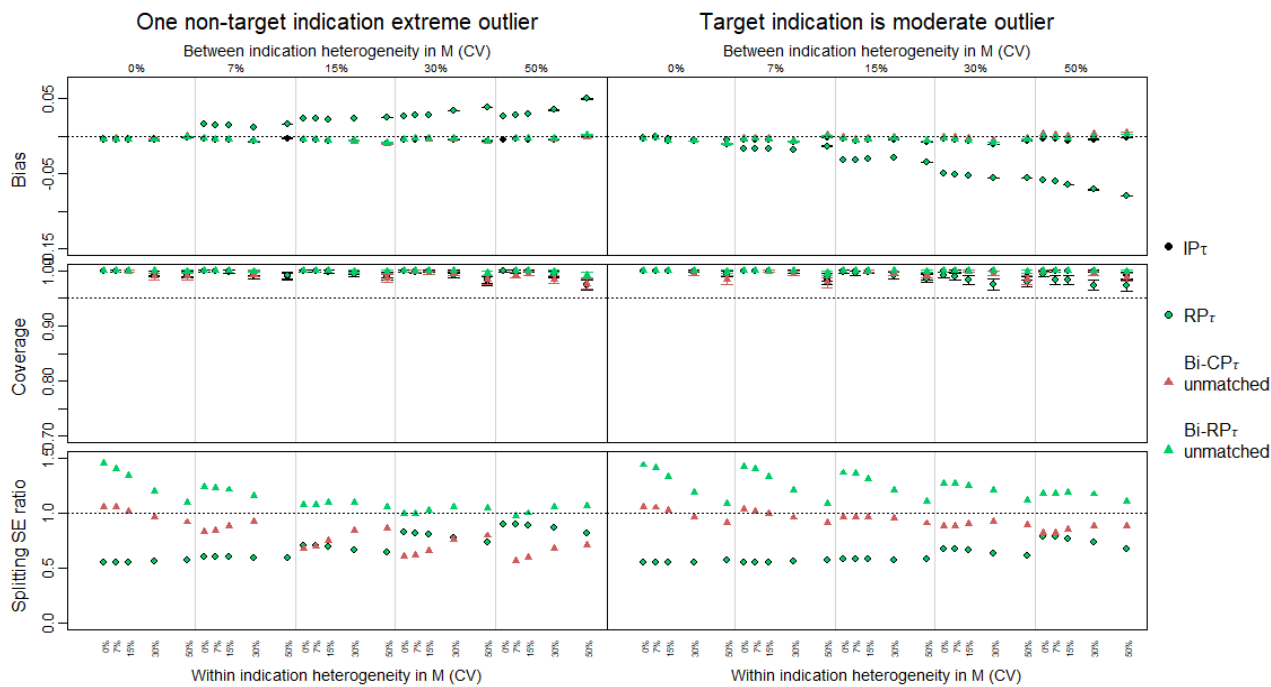

Figure 4: Performance measures comparing unmatched bivariate models (Bi-CP$_\tau$ and Bi-RP$_\tau$) and the univariate models IP$_\tau$ and RP$_\tau$ when there is OS in the target indication. Results are shown for large (Panel A) and small (Panel B) evidence base. The left panels show performance when the second largest indication was an outlier indication, the right panels show performance for when the target indication was an outlier indication and there is within and between indication heterogeneity in $\lambda_{01}$, $\lambda_{02}$ and $\Delta$. Results are shown for all combinations of within indication heterogeneity (CV 0%, 7%, 15%, 30%, 50%) and between indications (CV 0%, 7%, 15%, 30%, 50%). CV = coefficient of variation; CP = common parameters; RP = random parameters; unmatched = use an independent parameters model to estimate progression free survival; $\tau$ = common within indication heterogeneity; $\tau_j$ = independent within indication heterogeneity.

# 5. Discussion

This paper presents a simulation study exploring alternative multi-indication synthesis methods, considering a variety of data structures reflective of potential HTA contexts. This study focussed on evaluating the use of these methods in supporting predictions of the effect of a hypothetical treatment on OS in a "target" indication, by leveraging data from a broader evidence base across multiple indications. Although this study focuses on the impact of sharing information across indications, the results can be generalised to other sharing contexts, for example, sharing between drugs of the same class (Papanikos, Thompson et al. 2020).

We first explored simpler multi-indication synthesis methods that analyse LHR OS data directly (here termed univariate), including CP and RP models that respectively assume a common or exchangeable effect across indications. Our results showed that, when relative effectiveness across all indications was generated from the same process, with sharing being therefore plausible, these methods were unbiased and reduced uncertainty compared to using indication-specific evidence alone. This held even in the presence of heterogeneity between indications, although at higher levels of heterogeneity the borrowing of information is reduced. Simulated scenarios where this heterogeneity was low showed that CP sharing worked well, leading to appropriate descriptions of uncertainty and strong borrowing of information. Contexts of low heterogeneity align with the assumption of a common effect underlying CP synthesis models. When this heterogeneity was high, RP was more suitable, providing better calibrated estimates and showing that the assumption of exchangeability underlying RP estimates can accommodate the increased heterogeneity. However, this required larger datasets as when small datasets were considered (6 studies across 3 indications), RP was not well calibrated, generating higher uncertainty in predictions than was appropriate i.e. underconfidence. In any case, high heterogeneity lends itself to limited borrowing of information and limited increase in the precision of estimates. The value of sharing in this context is constrained by this.

We examined the case where one of the non-target indications was generated from a different, divergent, process (an outlier indication) and should not be shared from. Multi-indication synthesis (also applied on the outlier) posed a risk of bias, except where OS data for the target indication were available and an RP model was used for analysis (assuming a common within-indication heterogeneity parameter). Without OS data in the target indication, all univariate methods were demonstrated to be biased. Our simulations assumed mature data (80% of events observed) but in HTA practice, OS data is often less mature, so real world conclusions may fall between our "with OS" and "without OS" scenarios.

After looking at simpler univariate models, we examined whether mixture models could identify and correct for the outlier indication. Mixture models are argued to be more robust to outliers (Schmidli, Gsteiger et al. 2014, Neuenschwander, Wandel et al. 2016, Röver, Wandel et al. 2019). However, we found that, under the data structures typical in HTA contexts, these models were not able to identify the outlier and eliminate bias. The added complexity of these methods may therefore offer little value in practice for HTA.

We also explored the performance of bivariate (surrogacy) multi-indication synthesis models in predicting LHR OS effects for a target indication, under ideal conditions for surrogacy. When there is OS in the target indication, both univariate and unmatched bivariate models can correct for bias due to outlier indications. Due to the additional complexity of surrogacy models and the limited added value, they may not be useful in this case. However, the potential gains from surrogacy are clearer when i) the target indication itself is an outlier and ii) there are non-target outlier indications and no OS in the target indication (but there is PFS). In these cases, unmatched surrogacy models can provide low bias estimates for HTA decision making. This use of surrogacy has been considered elsewhere (Tai, Latimer et al. 2021, Wissinger, Koufopoulou et al. 2023).

Our simulation study used a mechanistic model (a three-state cancer progression MSM) to generate datasets, but analysed these assuming univariate or bivariate normality of the LHR for PFS and OS. The mechanistic element provides a novel way to examine the accuracy of the commonly applied synthesis models, emphasising that these may be misspecified. It is when analysing the relationship between LHR PFS and OS that we believe misspecification may have more marked influence from neglecting non-linearity. Further research should explore the reliable conditions for linear surrogate relationship and the impact of deviations from linearity in the accuracy of the bivariate models used for analyses here which are based on a linear surrogacy assumption. In examining the conditions for linear surrogacy, a critical consideration is the inclusion of heterogeneity in time to progression or pre-/ post-progression survival, which was not considered in our study but is realistic, as differences in prognosis are expected between indications (such as between prostate and breast cancer). (see Appendix A2). An alternative approach to imposing linear surrogate relationship is to fit a MSM model directly, however this is complex and uncommon in practice (Jansen, Incerti et al. 2023).

For multi-indication meta-analysis methods to be impactful in HTA, typical data structures must be considered. Multi-indication methods can provide most support for decision making when used early in the indication roll out when the number of approved indications are small. In HTA there are often relatively few indications, few studies per indication and immature evidence, especially for OS. This, combined with heterogeneity within and between indications, can account for the poor performance of highly parameterised synthesis models such as mixture models.

As with any simulation study, there are a number of ways to further extend and improve the generalisability of our findings, including extending the DGM to include heterogeneity in parameters other than the treatment effect, extending the scenarios evaluated, or considering alternative estimands more relevant for HTA decision making that consider, for example, the joint estimation of PFS and OS or immaturity in data.

Clinical judgement is crucial to assess whether sharing data from each indication is plausible and further research is required on how to gather and integrate these judgements. This could be in the form of selecting indications to include in the dataset or applying quantitative weights to indications based on their relevance to the target indication.

# 6. Conclusion

This analysis has implications for HTA decision making, which currently relies on single indication data to make decisions about multi-indication drugs. Our results showed that, for typical HTA contexts, univariate multi-indication methods can reduce uncertainty without inflating bias, particularly where there is OS data on the target indication. Mixture models add complexity and are unlikely to provide improvements in practice, so should not be used. In cases where univariate models may become biased, such as with outlier target indications, bivariate surrogacy models show potential but further research is necessary to understand their performance under more realistic conditions.

# Acknowledgements

This work was funded by a Medical Research Council (MRC) better methods better research grant MR/W021102/1. DG received the financial support of the European Union's Horizon Europe Excellent Science programme under the Marie Skłodowska-Curie Actions Grant Agreement [Grant Agreement No 101081457] and in part from Research Ireland under grant number 13/RC/2073_P2.

The Viking cluster was used during this project, which is a high-performance computer facility provided by the University of York. We are grateful for computational support from the University of York, IT Services and the Research IT team.

# A1. Estimating multi-state parameters based on median PFS and OS

In this section we describe our approach to find reasonable multi state model parameters to ground our simulation study. We do this using the bevacizumab case study described in Singh et al (2023). Specifically, we want to find values for $\lambda_{01}, \lambda_{02}$ and $\Delta$ which represent their mean value across the indications in the bevacizumab case study. As described in the paper mean value of M was assumed to 0.6 ($\mu_M$ = -0.511 on log scale) and a wide range of values of within and between heterogeneity were explored.

A "initial-progressed-dead" multistate model equivalent to the one used in the main paper is illustrated below. This differs from the main paper in that it does not include a treatment effect on progression (M) and the rate of post progression death is parameterised using $\lambda_{12}$ rather than as $\lambda_{12}.\Delta$.

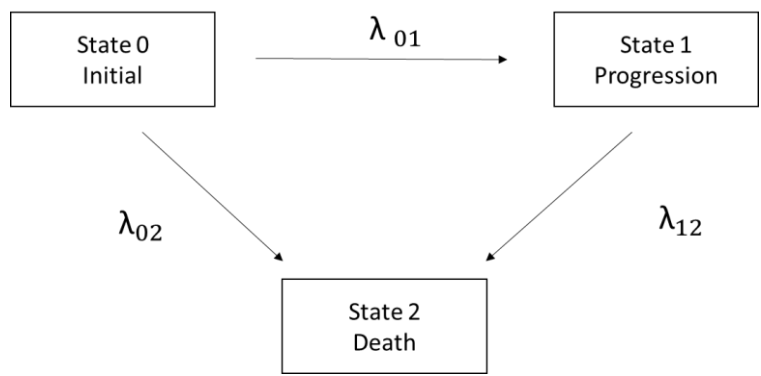

*Figure A1.1: "Initial-progressed-dead" multistate model as described in Erdmann et al (2025).*

With the objective of estimating transition rates for the model shown in Figure A1.1, median OS and PFS were extracted from the control arms of the larger studies within each cancer type (among those studies that reported both PFS and OS), see Table A1.1.

| Cancer type | Publication | Line of treatment | Median, in months | | $\lambda_{01}^{ctrl}$ | $\lambda_{02}^{ctrl}$ | $\lambda_{12}^{ctrl}$ | $\Delta$ |
|---|---|---|---|---|---|---|---|---|
| | | | PFS ctrl | OS ctrl | | | | |
| BRE | Brufsky (2011) | Second | 5.1 | 16.4 | 0.126 | 0.0102 | 0.072 | 7.0588 |
| BRE | Cameron (2008) | First | 5.8 | 24.8 | 0.109 | 0.0102 | 0.039 | 3.8235 |
| BRE | Miles (2013) | First | 8 | 31.9 | 0.076 | 0.0102 | 0.03 | 2.9412 |
| CER | Tewari (2017) | First | 6 | 13.3 | 0.105 | 0.0102 | 0.124 | 12.1569 |
| COL | Saltz (2008) | First | 8 | 19.9 | 0.076 | 0.0102 | 0.07 | 6.8627 |
| COL | Schmoll (2012) | First | 9.9 | 22.8 | 0.06 | 0.0102 | 0.064 | 6.2745 |
| COL | Bennouna (2013) | Second | 4.1 | 9.8 | 0.159 | 0.0102 | 0.161 | 15.7843 |
| GLIO | Chinot (2014) | First | 6.2 | 16.7 | 0.102 | 0.0102 | 0.08 | 7.8431 |
| GLIO | Wick (2017) | Second | 1.5 | 8.6 | 0.452 | 0.0102 | 0.107 | 10.4902 |
| GLIO | Gilbert (2014) | First | 7.3 | 15.7 | 0.085 | 0.0102 | 0.106 | 10.3922 |
| NSCLC | Sandler (2005) | First | 4.5 | 10.3 | 0.144 | 0.0102 | 0.16 | 15.6863 |
| NSCLC | Reck (2019) | First | 6.9 | 21.4 | 0.09 | 0.0102 | 0.054 | 5.2941 |

| | | | | | | | |
|---|---|---|---|---|---|---|---|
| NSCLC | Zhou (2015) | First | 6.5 | 17.7 | 0.096 | 0.0102 | 0.075 | 7.3529 |
| OFTPP | Oza (2015) | First | 17.5 | 58.6 | 0.029 | 0.0102 | 0.013 | 1.2745 |
| OFTPP | Aghajanian (2012) | Second | 12.4 | 35.2 | 0.046 | 0.0102 | 0.031 | 3.0392 |
| OFTPP | Burger (2011) | First | 10.3 | 39.3 | 0.057 | 0.0102 | 0.023 | 2.2549 |
| REN | Escudier (2007) | First | 5.4 | 19.8 | 0.118 | 0.0102 | 0.054 | 5.2941 |

*Table A1.1: This table shows the median PFS and OS values extracted from control arms of larger studies in the bevacizumab case study reported in Singh et al (2023). It also shows the multi state model parameters assumed or derived from these studies. OS = overall survival; PFS = progression free survival; BRE = breast; CER = cervical; COL = colorectal; GLIO = glioblastoma; NSCLC = non-small cell lung cancer; OFTPP = Ovarian, fallopian tube and primary peritoneal; REN = renal.*

To derive the multi-state parameters from the extracted median PFS results, we assume that PFS is a random variable which follows an exponential distribution. Therefore, median(PFS)=$\frac{log(2)}{\lambda}$, where λ is the rate parameter. In the case of the control arm of the multistate model λ = $\lambda_{01}^{ctrl} + \lambda_{02}^{ctrl}$. This results in the following formula:

$$\lambda_{01}^{ctrl} = \ln(2)/\text{median(PFS)} - \lambda_{02}^{ctrl}.$$

From first-line treatment of adult patients with NSCLC in Jansen et al (2023) $\lambda_{02}^{ctrl}$ = 0.0102. This was assumed constant across all studies. By plugging in $\lambda_{02}^{ctrl}$ and the observed control arm median(PFS) from Table A1.1 above, we estimated $\lambda_{01}^{ctrl}$ for each study.

Given exponential transitions, Erdmann et al (Erdmann, Beyersmann et al. 2025) provide equations which deterministically relate the multi-state model parameters in Figure A1.1 ($\lambda_{01}, \lambda_{02}, \lambda_{12}$) to survival functions (and therefore medians, S(t) = 0.5) for PFS and OS. For each trial we used these equations and the values for $\lambda_{01}^{ctrl}, \lambda_{02}^{ctrl}$ to back solve for the value of $\lambda_{12}^{ctrl}$ which is consistent with the median(OS) observed in the control arm of the trial. This was done using single parameter optimization in R.

The multi-state model described in the main text is defined in terms of $\lambda_{01}, \lambda_{02}$, Δ and M. Differences between the arms are defined using M so $\lambda_{01} = \lambda_{01}^{trt}$. Pre-progression mortality was assumed constant across treatment arms so $\lambda_{02} = \lambda_{02}^{ctrl}$. Δ was defined as a multiplier for the change in mortality post progression, it was assumed common across arms so is defined as $\lambda_{12}^{ctrl}/\lambda_{02}$.

Because $\lambda_{02}$ was assumed to be constant at 0.0102, the mean value across indications is μ$_{λ02}$ = 0.0102. To calculate the mean value of $\lambda_{01}$ and Δ across indications we need to first convert the parameters of interest in each study to the log scale. Then for each parameter we calculate the mean within each indication. Then, weighting each indication equally, we take the average of the indication means. This results in μ$_{λ01}$ = 0.097 and μ$_Δ$ = 6.32 on the natural scale.

# A2. Linear surrogacy between log hazard ratios for progression free survival and overall survival

As shown in Erdmann et al (Erdmann, Beyersmann et al. 2025), under the conditions of the MSM described in the main paper, LHR OS and LHR PFS are a function of the MSM parameters and duration of follow-up (t). The LHR for PFS is given by:

$$\text{LHR PFS} = \ln\left(\frac{\lambda_{01}.M + \lambda_{02}}{\lambda_{01} + \lambda_{02}}\right).$$

The OS hazard ($h_{OS}$) in the control and treatment groups are below:

$$h_{OS,ctrl} = \frac{\Delta(\lambda_{01} + \lambda_{02}) - \lambda_{01}.\lambda_{02} + \Delta.exp((\lambda_{01} - \Delta).t)}{\Delta - \lambda_{01}.exp((\lambda_{01} - \Delta).t)},$$

$$h_{OS,trt} = \frac{\Delta(\lambda_{01}.M + \lambda_{02}) - \lambda_{01}.M.\lambda_{02} + \Delta.exp((\lambda_{01}.M - \Delta).t)}{\Delta - \lambda_{01}.M.exp((\lambda_{01}.M - \Delta).t)}.$$

Therefore, the LHR OS is the log of the ratio of these hazards.

$$\text{LHR OS} = \ln\left(\frac{h_{OS,trt}}{h_{OS,ctrl}}\right).$$

To understand when approximately linear surrogacy is possible, we simulated from the equations above. We begin with the mean parameters used in the case study (and based on the bevacizumab dataset): $\lambda_{01}$ = 0.097, $\lambda_{02}$ = 0.01, Δ = 6.32, M = 0.6 on natural scale. For each MSM parameter we explore a wide range of values, investigating the impact on LHR PFS, LHR OS and their joint distribution (i.e. the surrogacy relationship). Note that in using the above equations we are simulating the true underlying study level values with no sampling uncertainty.

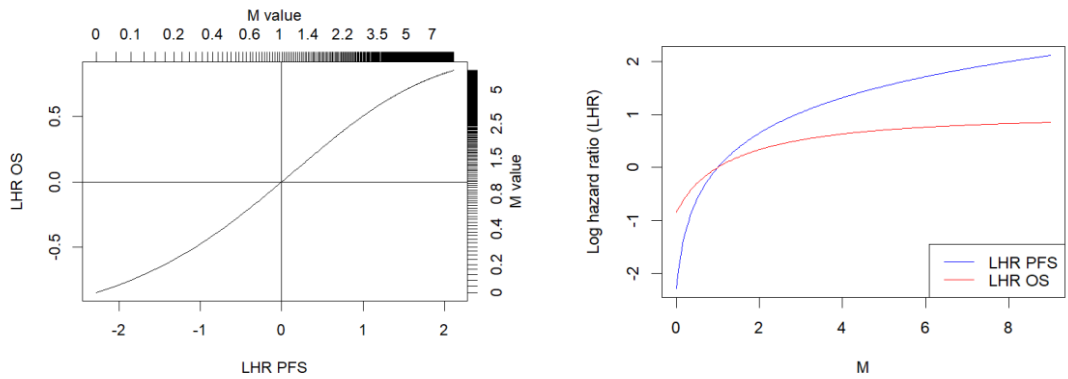

Above M is varied from 0.01 to 9 (natural scale). Surrogacy is approximately linear when the treatments have a similar effect i.e. M close to 1.

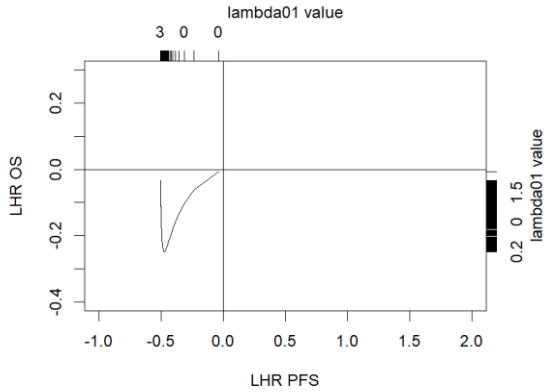
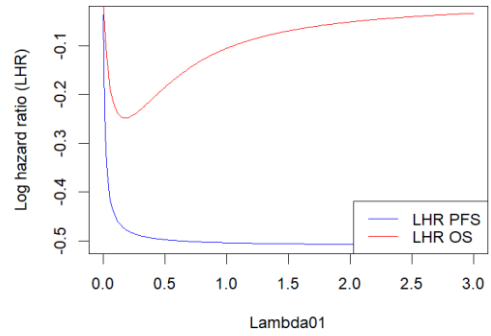

Above $\lambda_{01}$ is varied from 0.001 to 3 (natural scale). The surrogacy relationship is non-monotonic. This is because LHR OS is non-monotonic in $\lambda_{01}$ (see right hand graph).

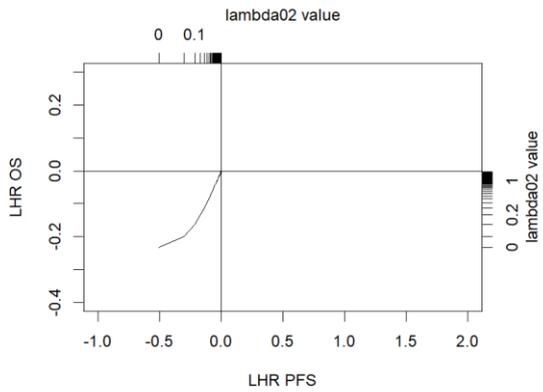
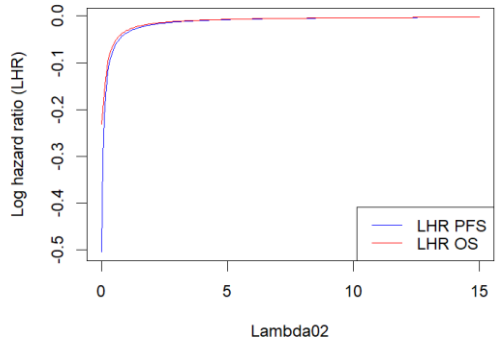

Above $\lambda_{02}$ ranges from 0.001 to 15 (natural scale). There is an approximately linear component between LHR PFS = 0 and -0.25.

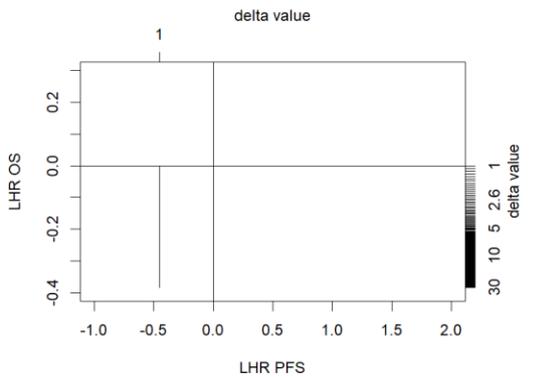
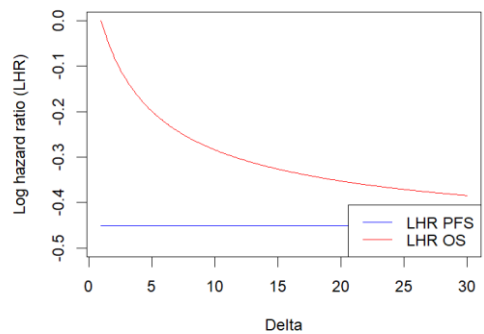

Above Δ ranges from 1 to 30 (natural scale). Because this is the increase in the rate of mortality post progression it has no impact on PFS. This mean that there is no surrogacy relationship induced by variation in Δ.

This analysis indicates that variation in $\lambda_{01}$, $\lambda_{02}$ and M can be associated with approximately linear surrogacy regions. In each case this region does not extend over all values of the MSM parameter. This indicates that linear surrogacy cannot be assumed generally.

Note that because the MSM is non-linear, the results above depend on the mean values of all the parameters in the model so any generalisation is challenging.

# A3. Definitions of model convergence

**Option 1: $\hat{R}$ < 1.1 for all parameters**

Fit model using JAGS then calculate $\hat{R}$ statistic for all model parameters. The model has converged if $\hat{R}$ < 1.1 for all parameters. Note that for mixture models "c" and "if branch" are included in the MCMC output but are indicators and not model parameters so $\hat{R}$ is not calculated for these model components.

**Option 2: $\hat{R}$ < 1.1 for parameters used in prediction**

The principle behind this approach is that convergence only matters in those parameters which are ultimately used to make the prediction for the estimand of interest. For example, the mixture models include some parameters which will not be updated if the model considers them implausible, so it does not make sense to check the convergence of these parameters. To implement this option, we need to define the parameters used in prediction for each model, this information is included in Table 1 of the main paper.

# A4. All simulation study results

# A4.1 Outlier indications

## A4.1.1 With overall survival in target indication

### Univariate non-mixture

**Large dataset**

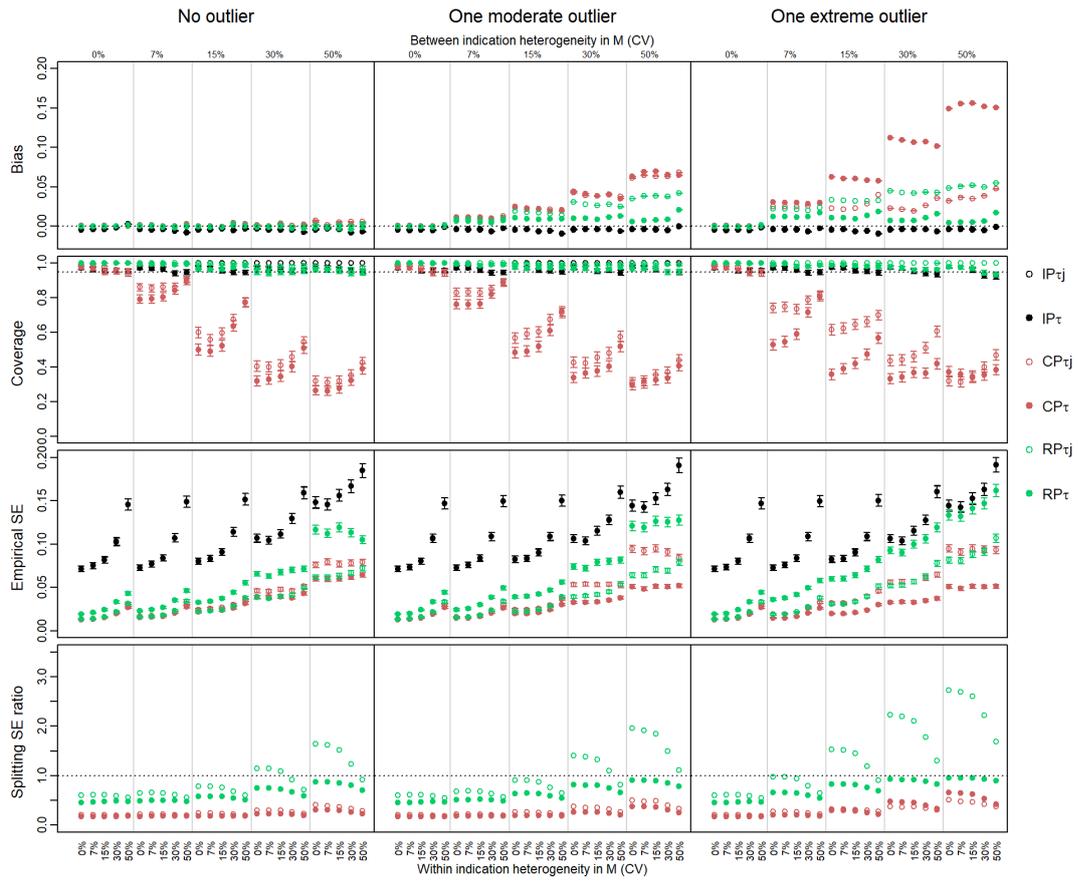

**Medium dataset**

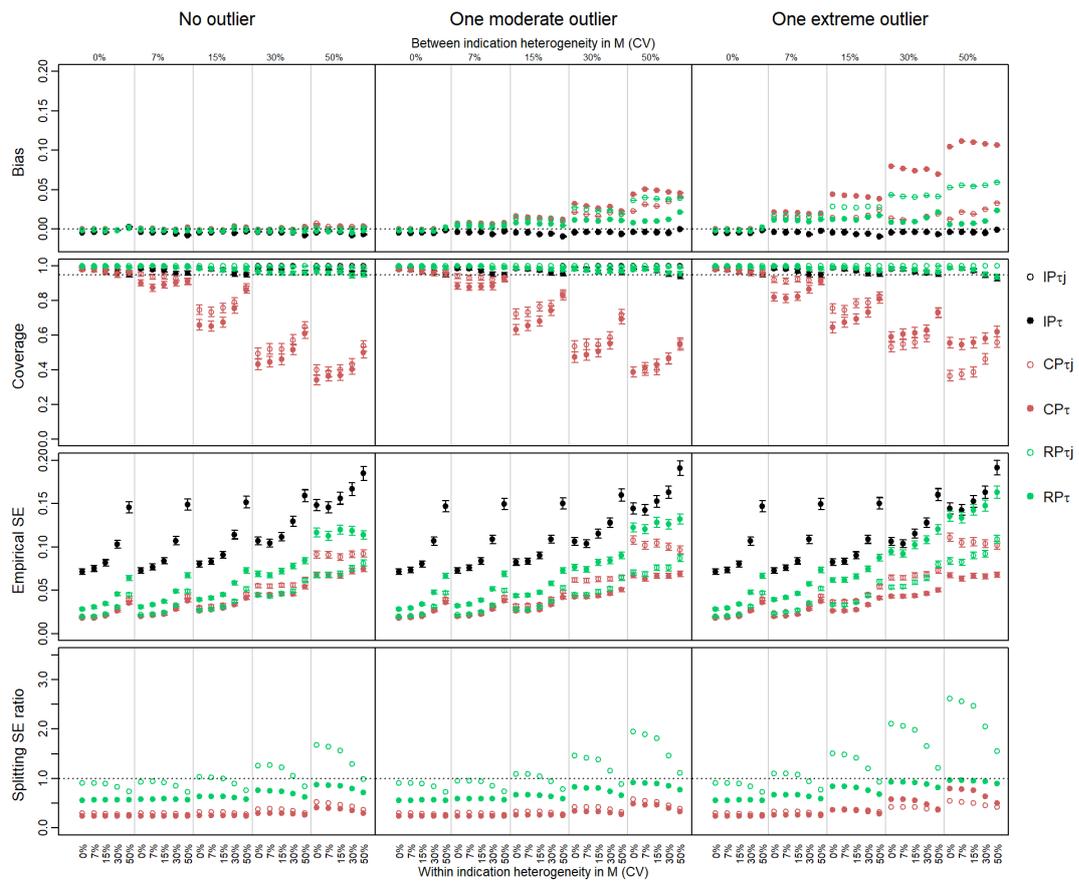

**Small dataset**

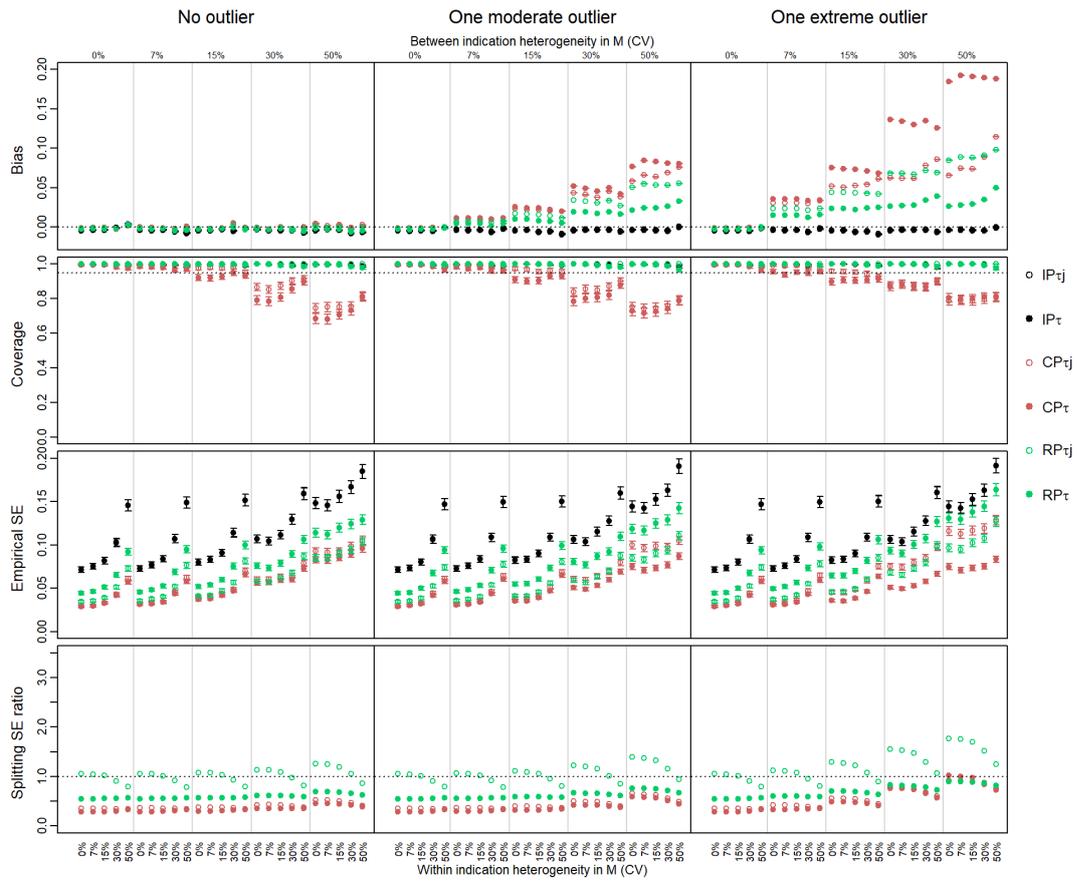

# Univariate mixture

## Large dataset

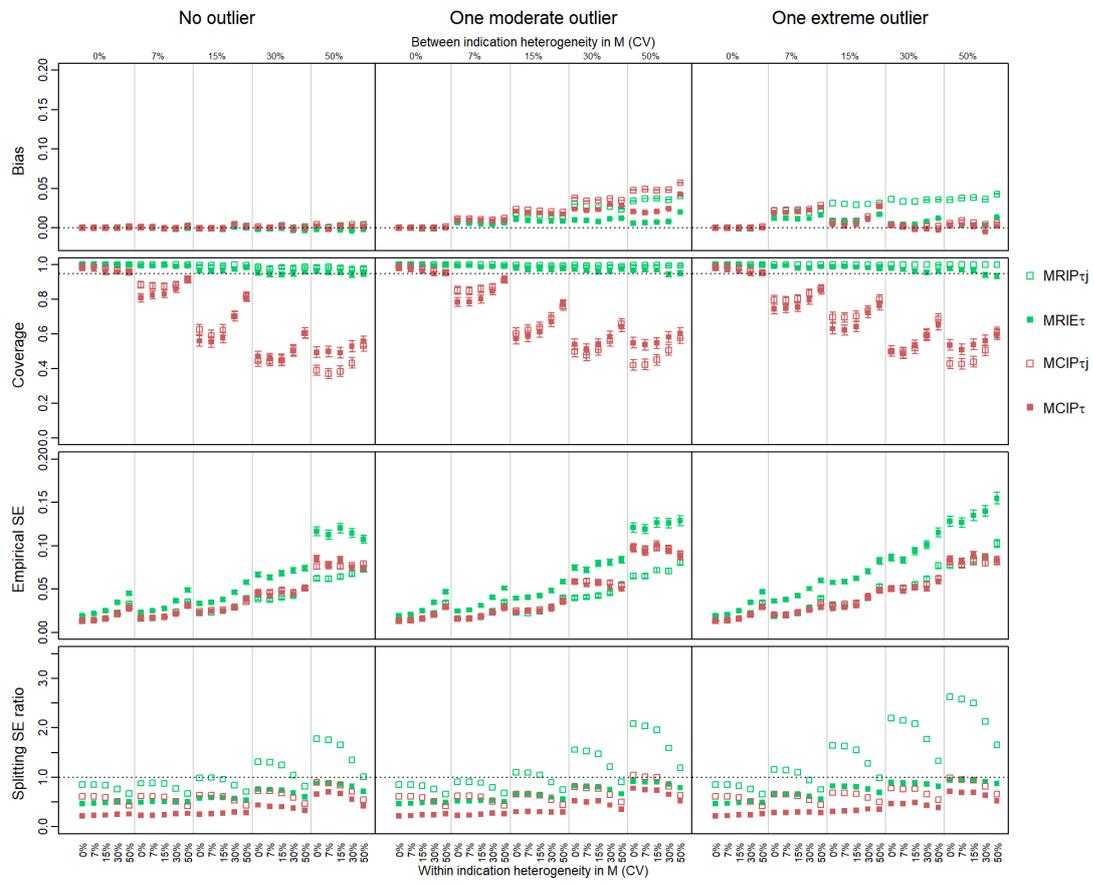

**Medium dataset**

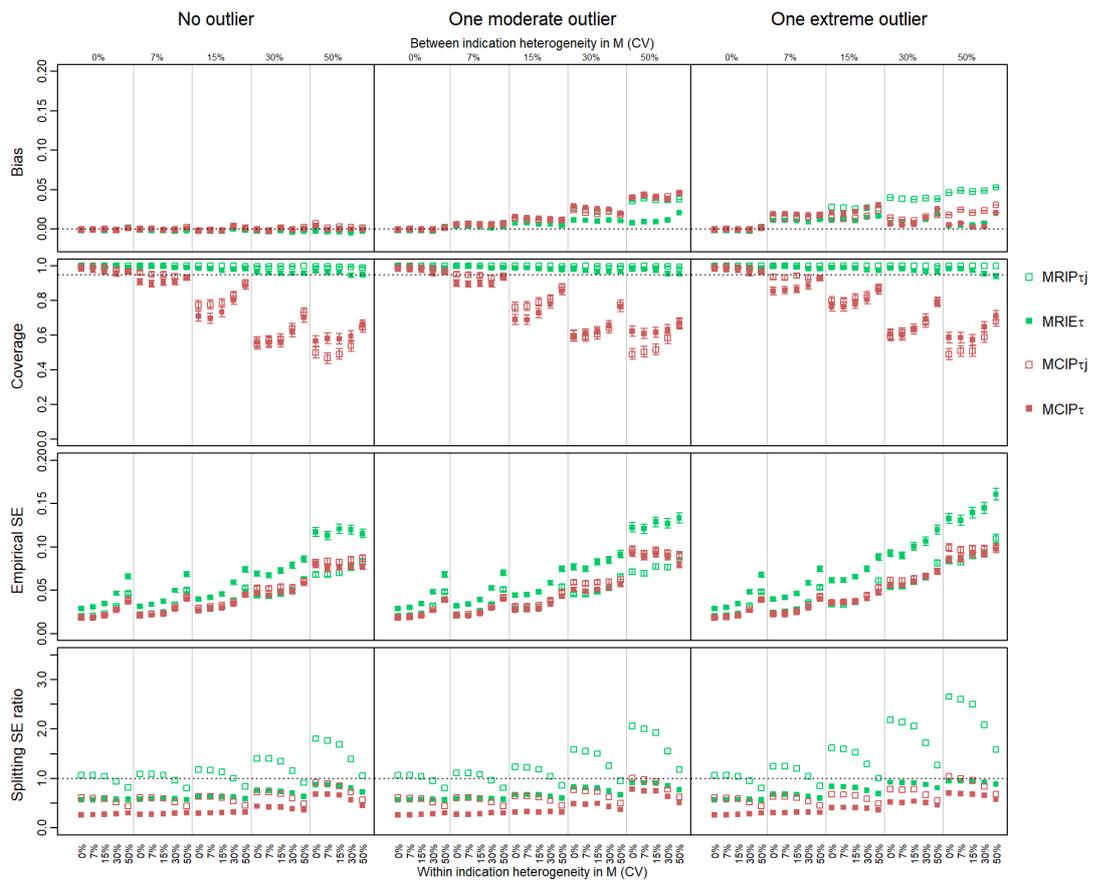

**Small dataset**

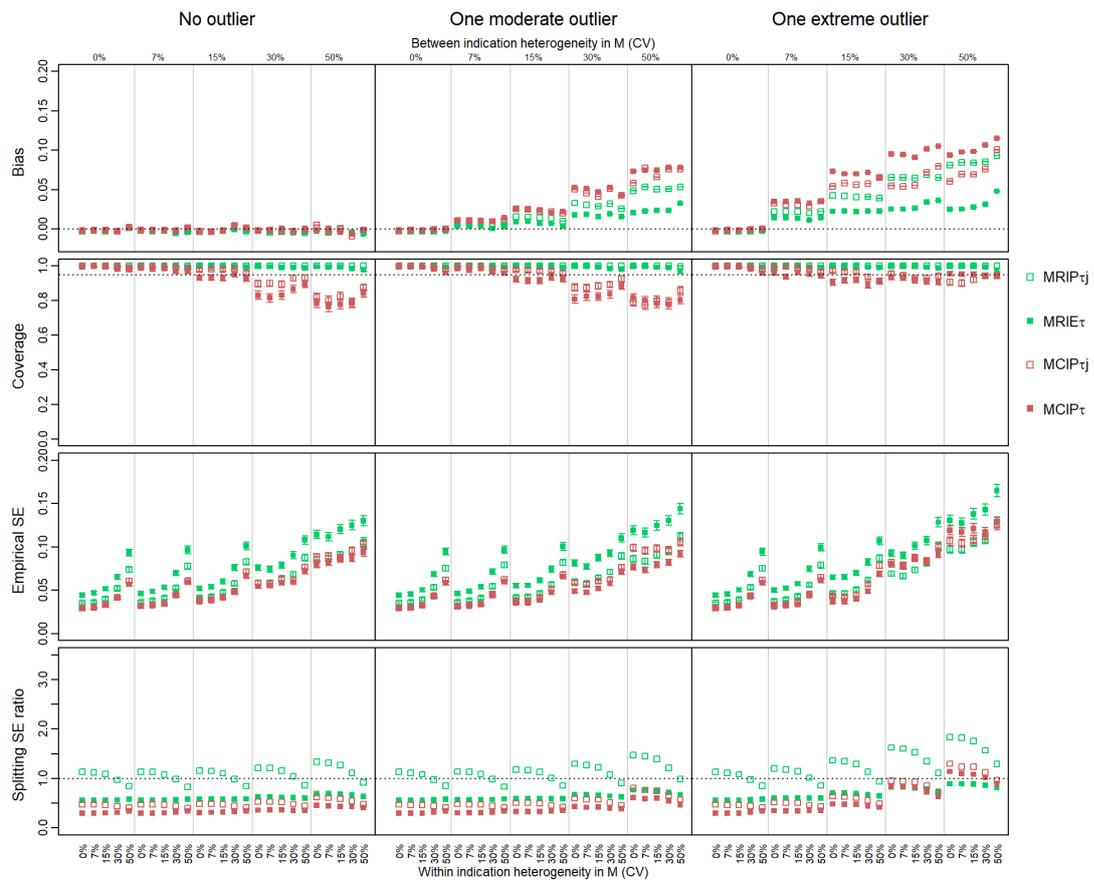

# Surrogate unmatched

## Large dataset

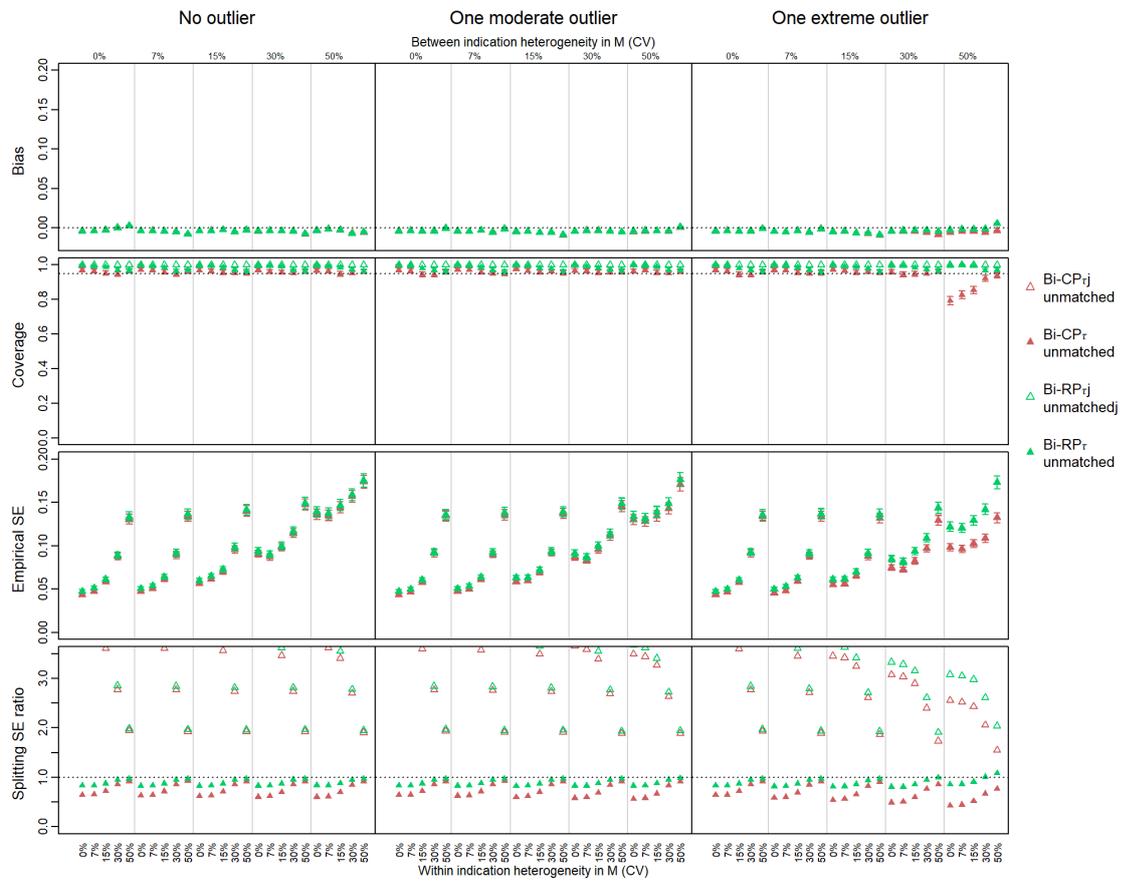

# Medium dataset

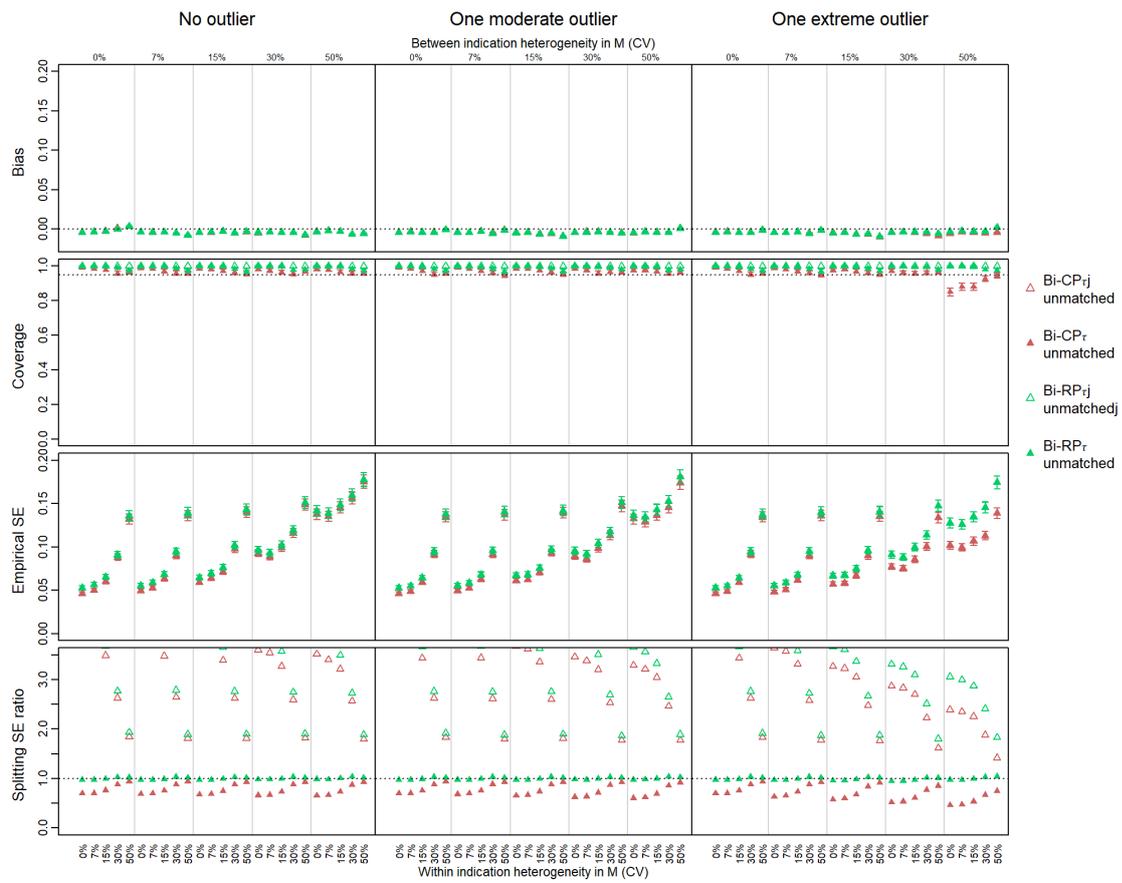

# Small dataset

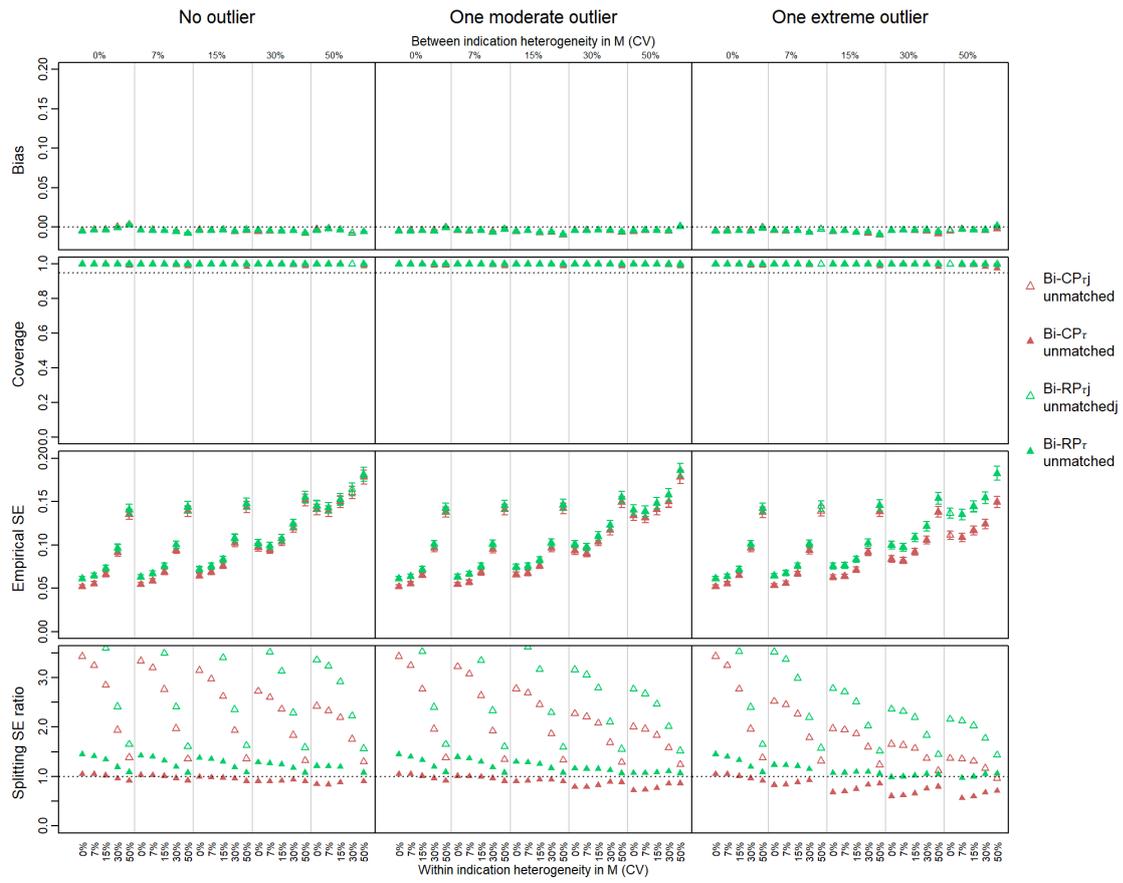

# Surrogate matched

## Large dataset

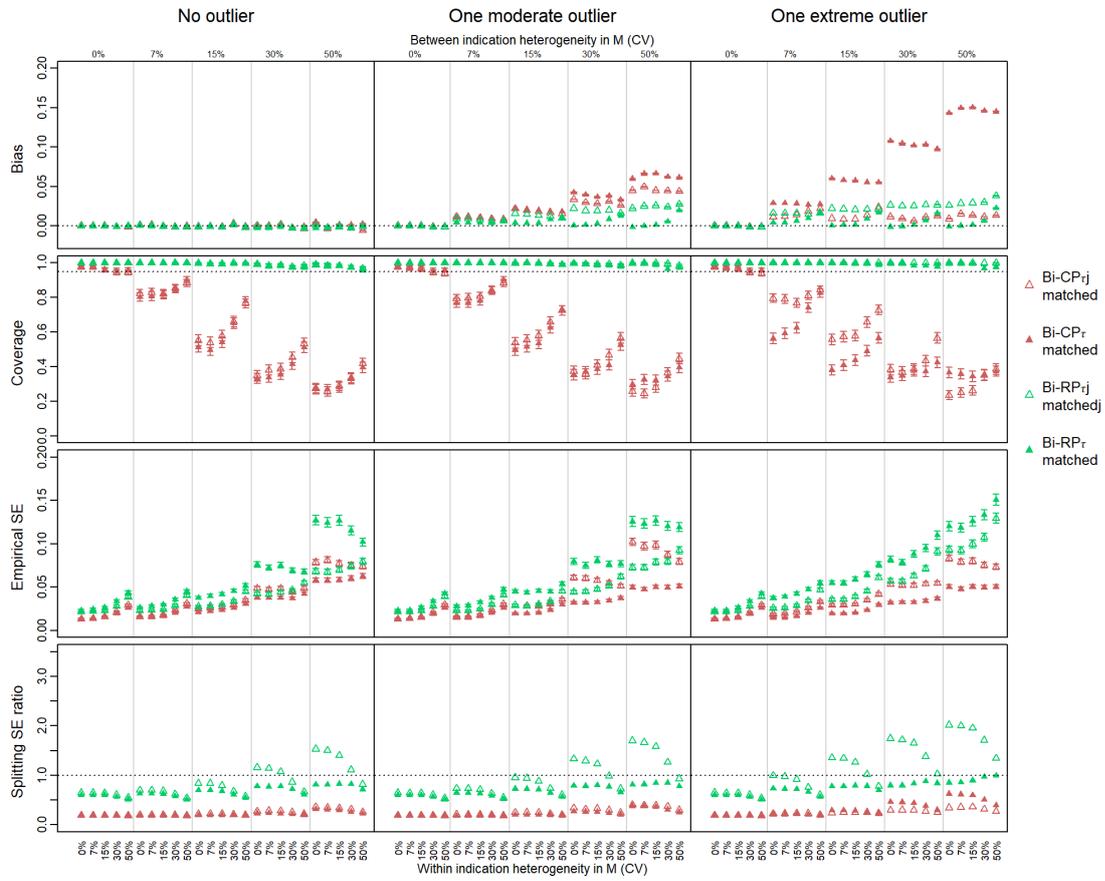

**Medium dataset**

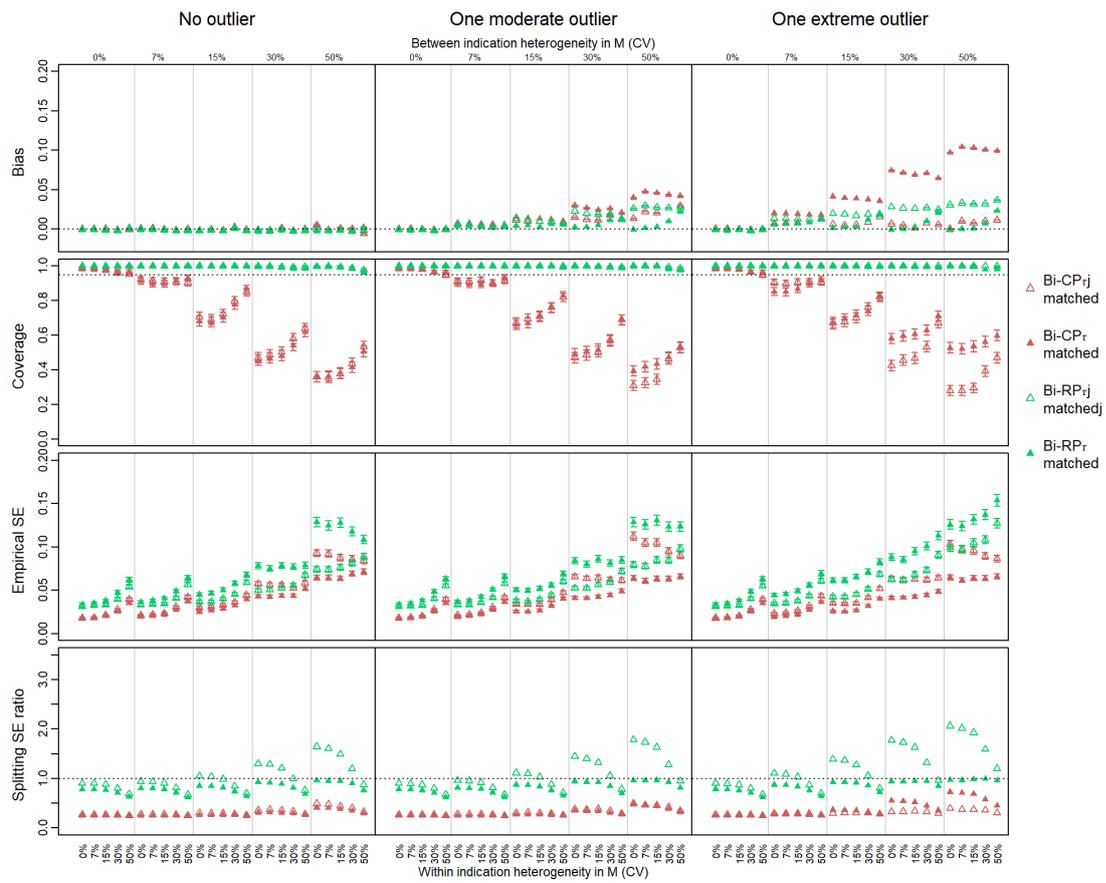

# Small dataset

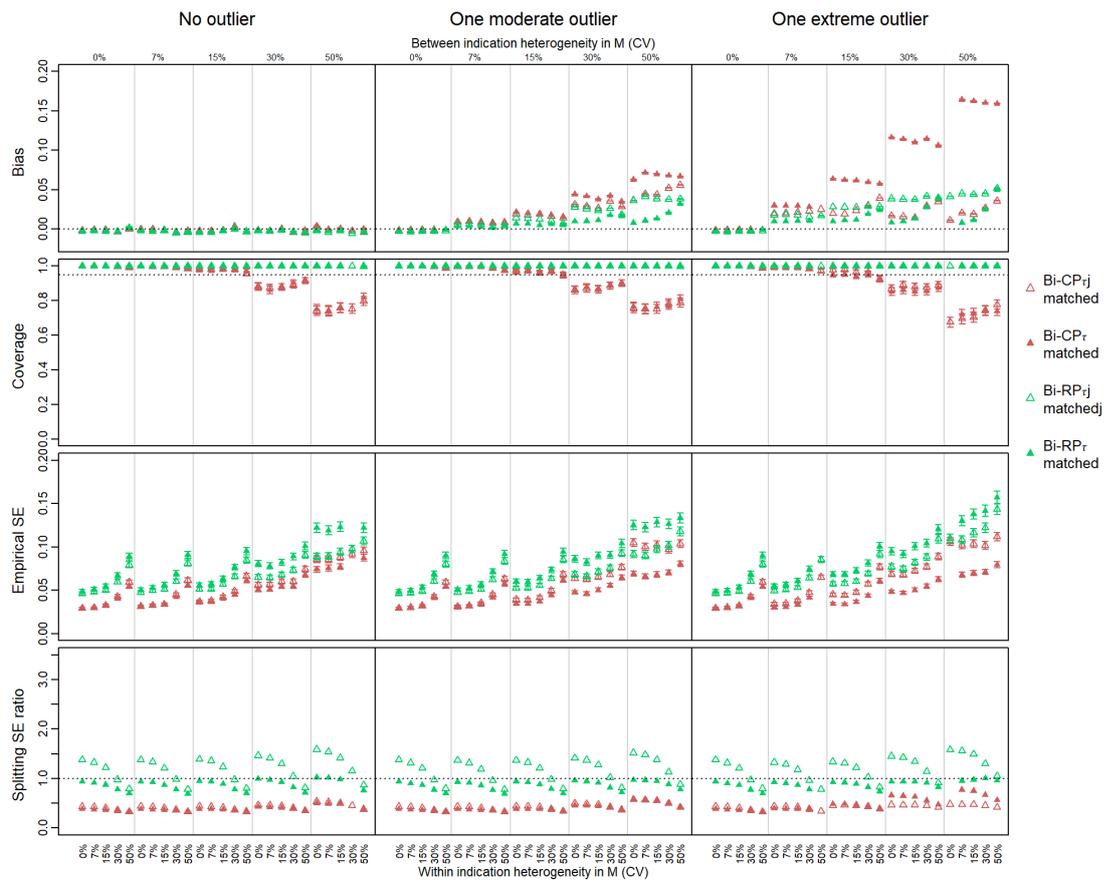

## A4.1.2 No overall survival in target indication

Note that because there is no OS in the target indication it is not possible to esimtate the univariate IP model. This causes issues with the interpretation of the splitting SE ratio because this is based on the IP model in the target indication. In the graphs below the splitting SE ratio is calculated comparing the SE of the sharing models to the SE of the IP model if there was OS in the target indication.

### Univariate non-mixture

**Large dataset**

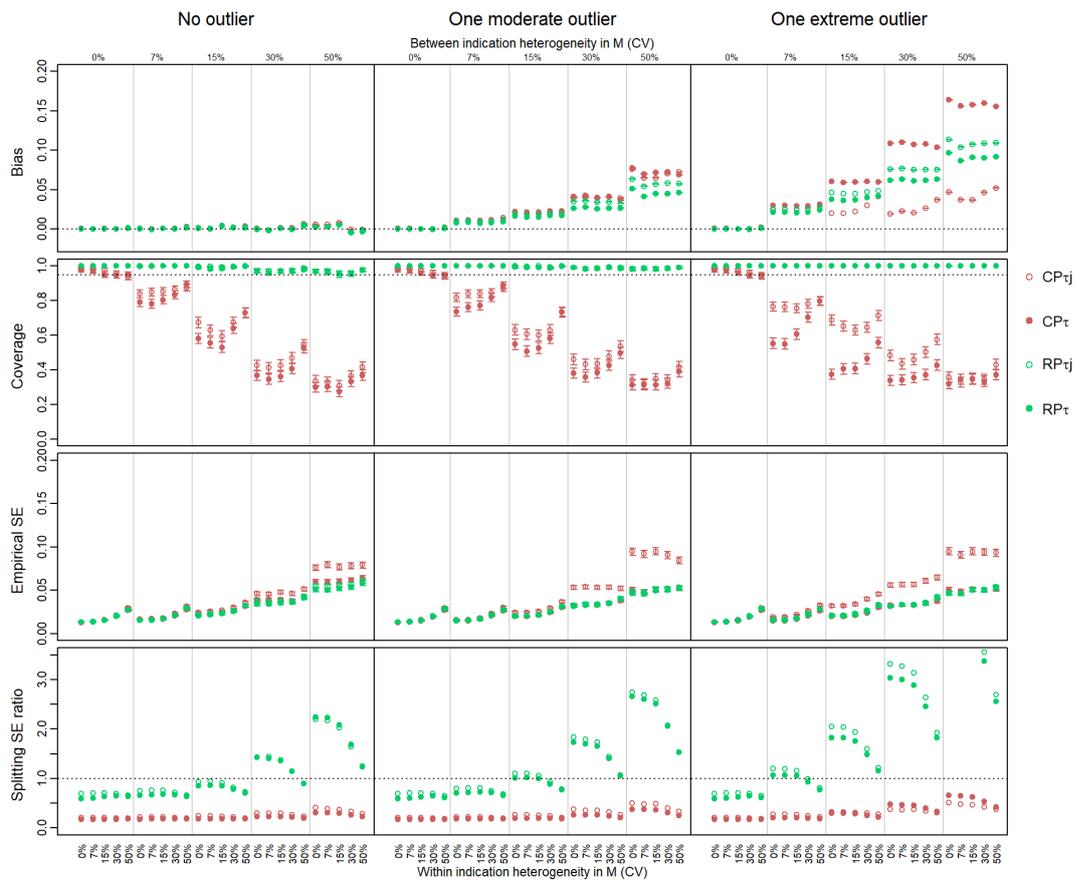

**Medium dataset**

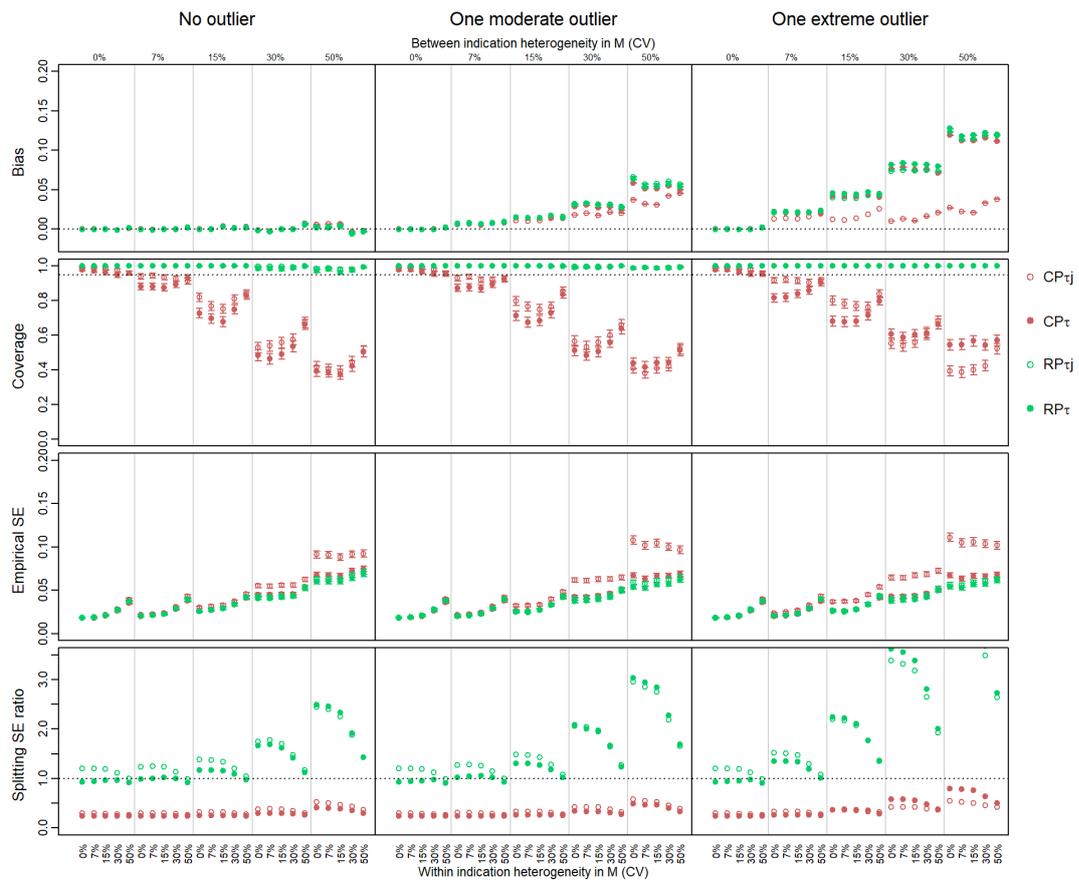

**Small dataset**

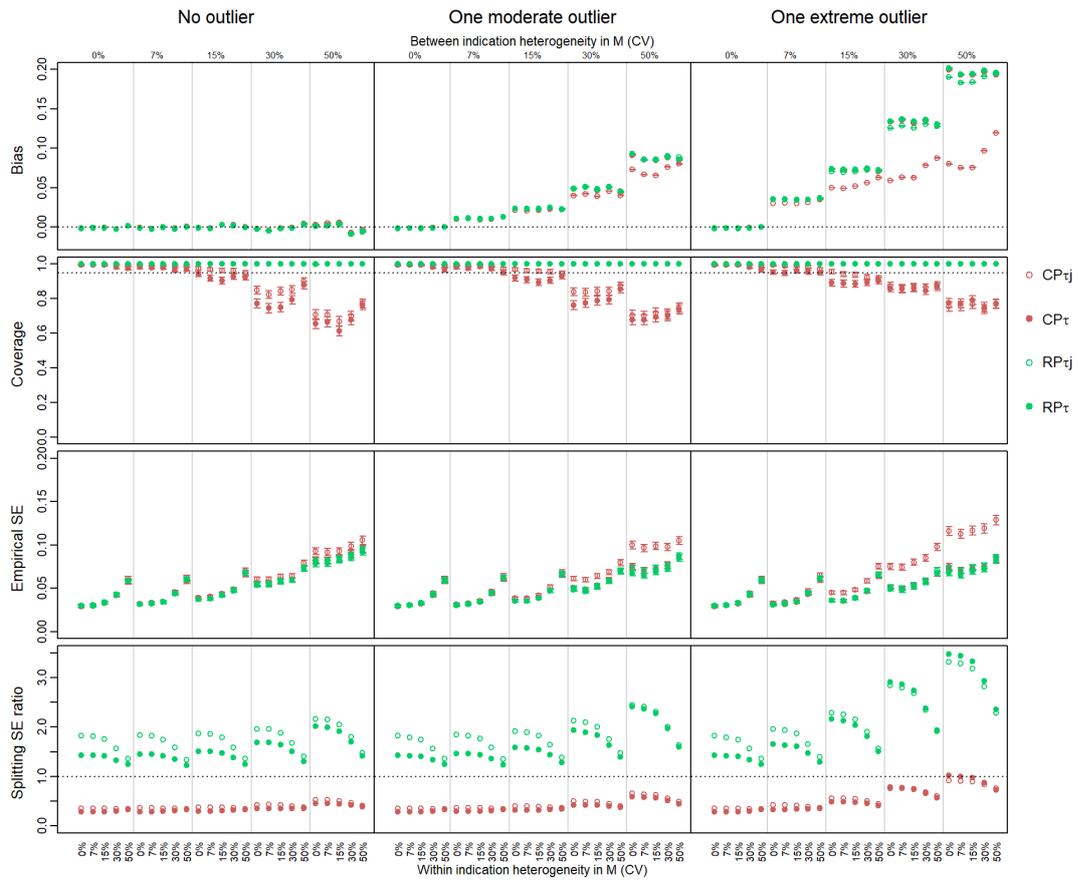

# Univariate mixture

**Large dataset**

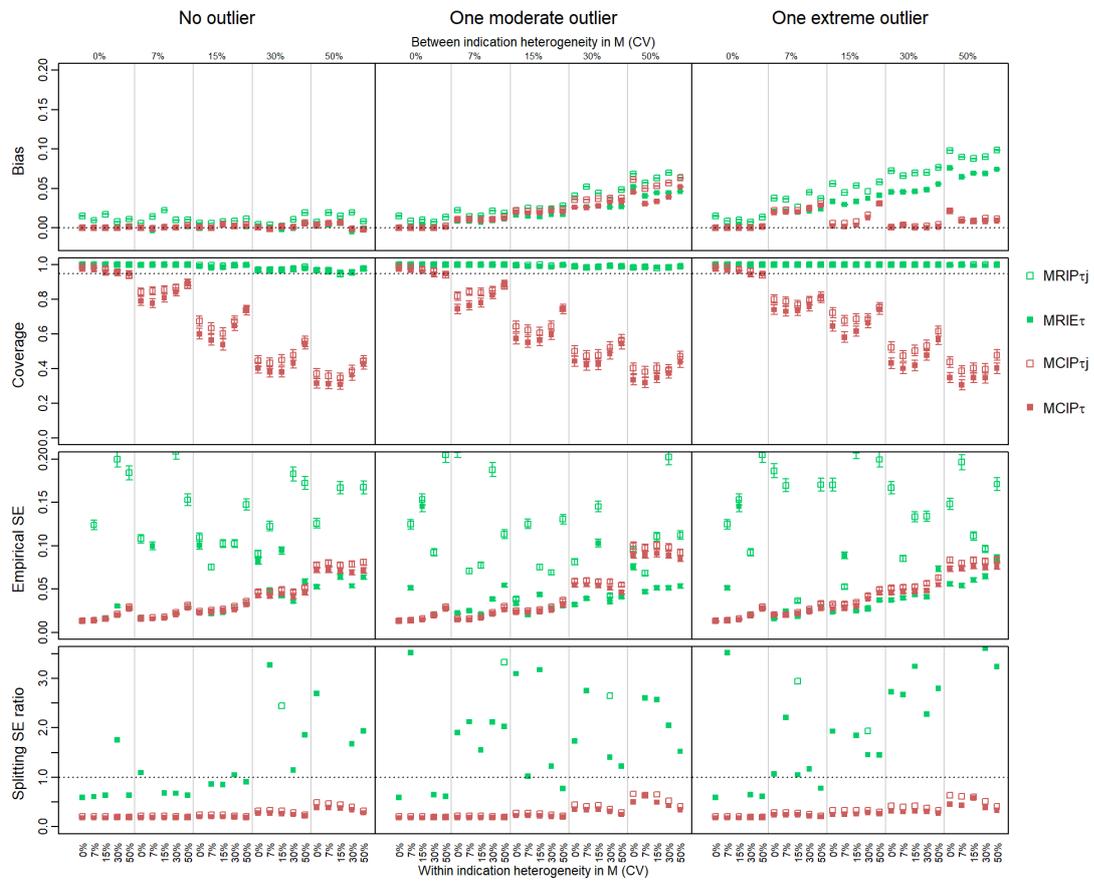

**Medium dataset**

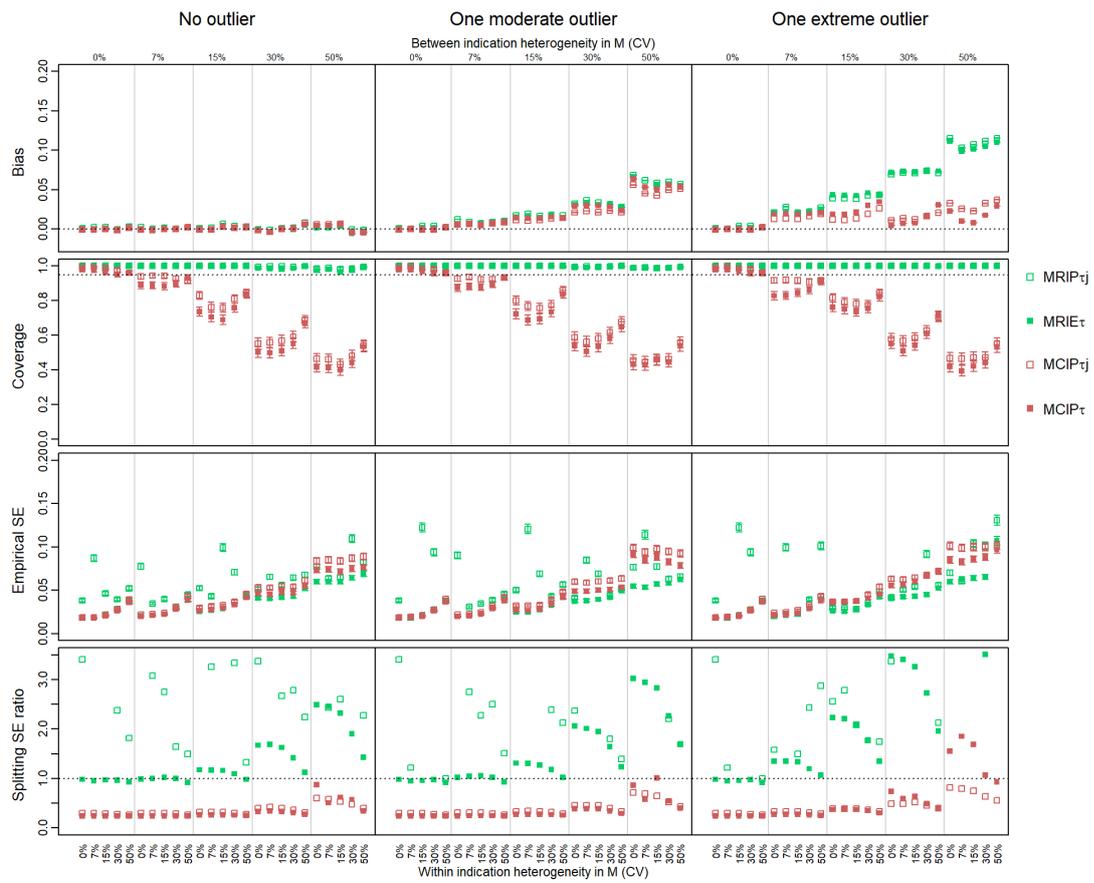

## Small dataset

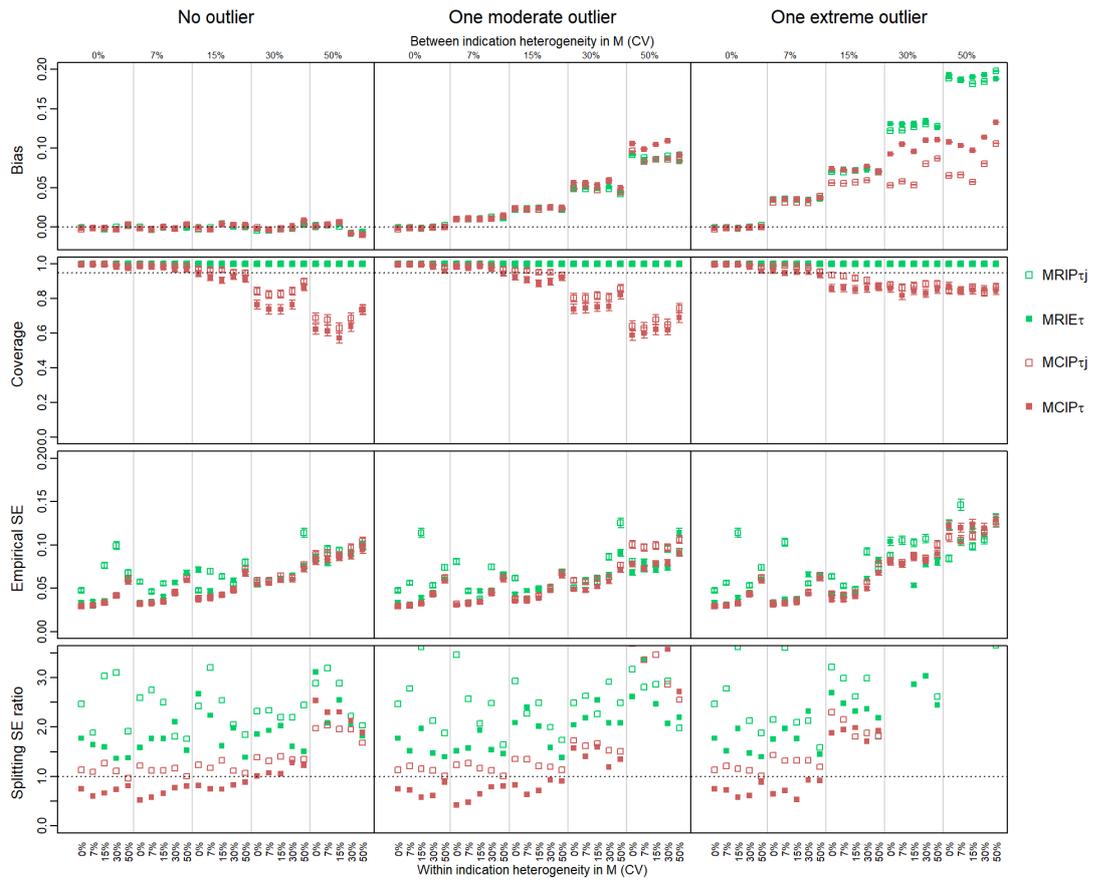

# Surrogate unmatched

## Large dataset

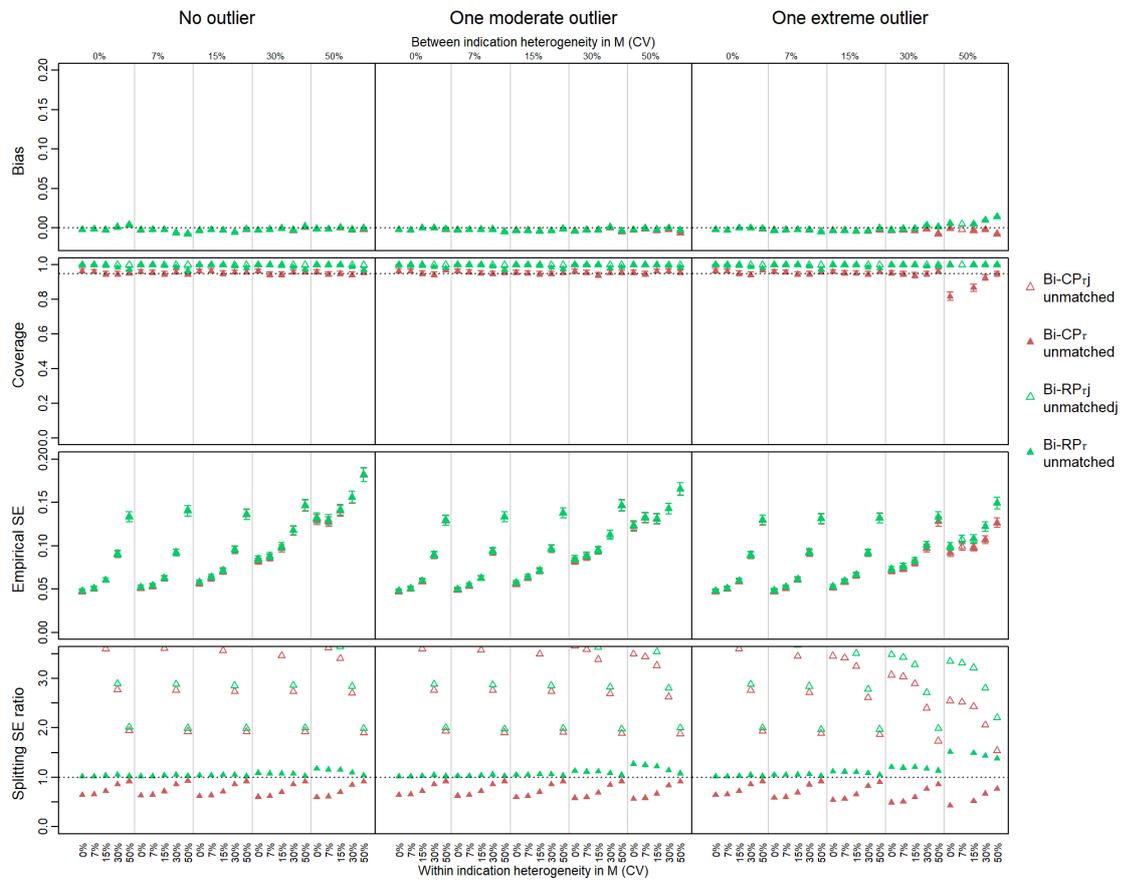

**Medium dataset**

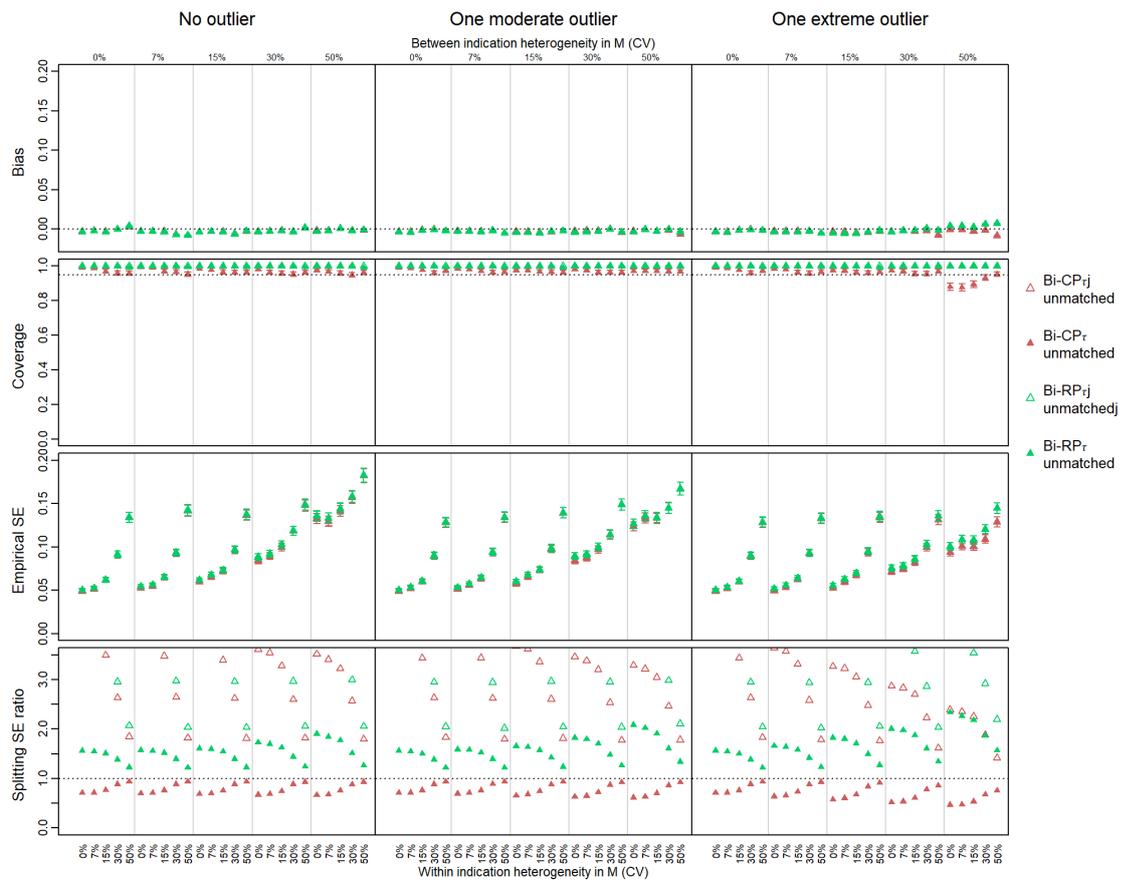

**Small dataset**

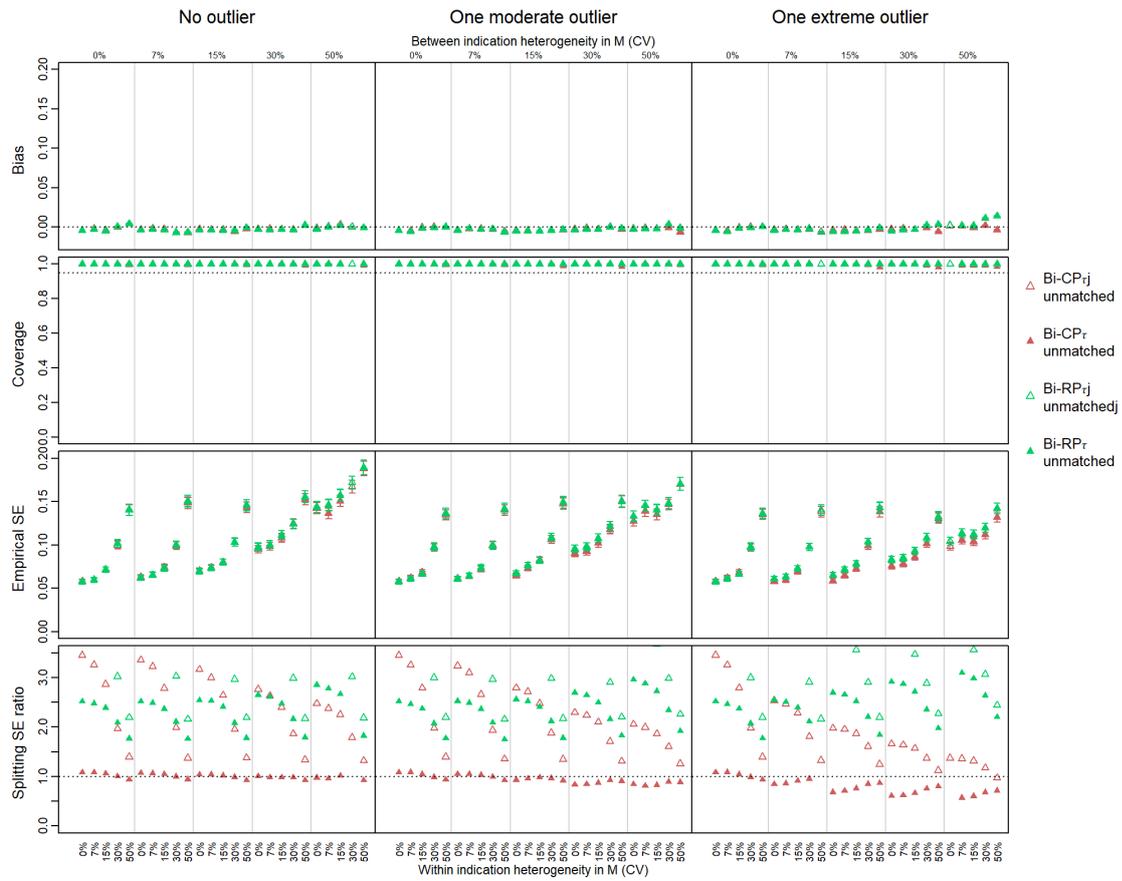

## Surrogate matched

### Large dataset

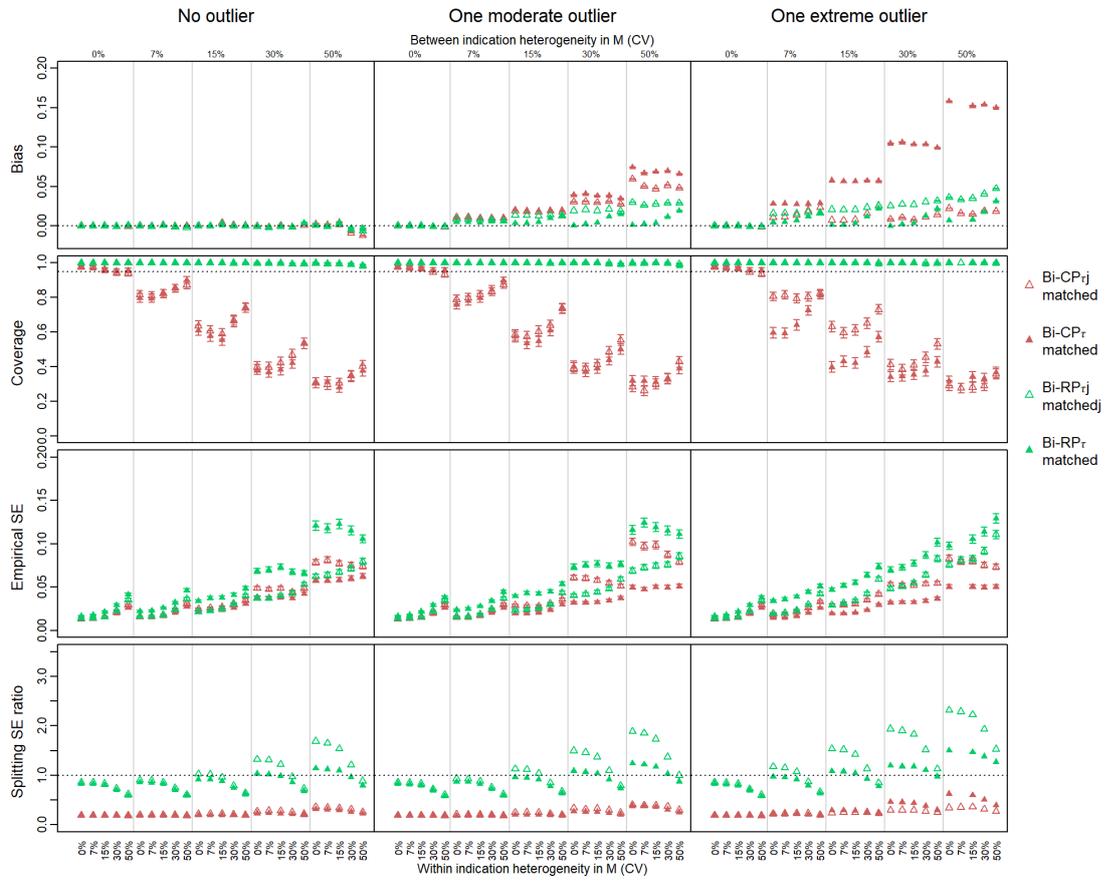

**Medium dataset**

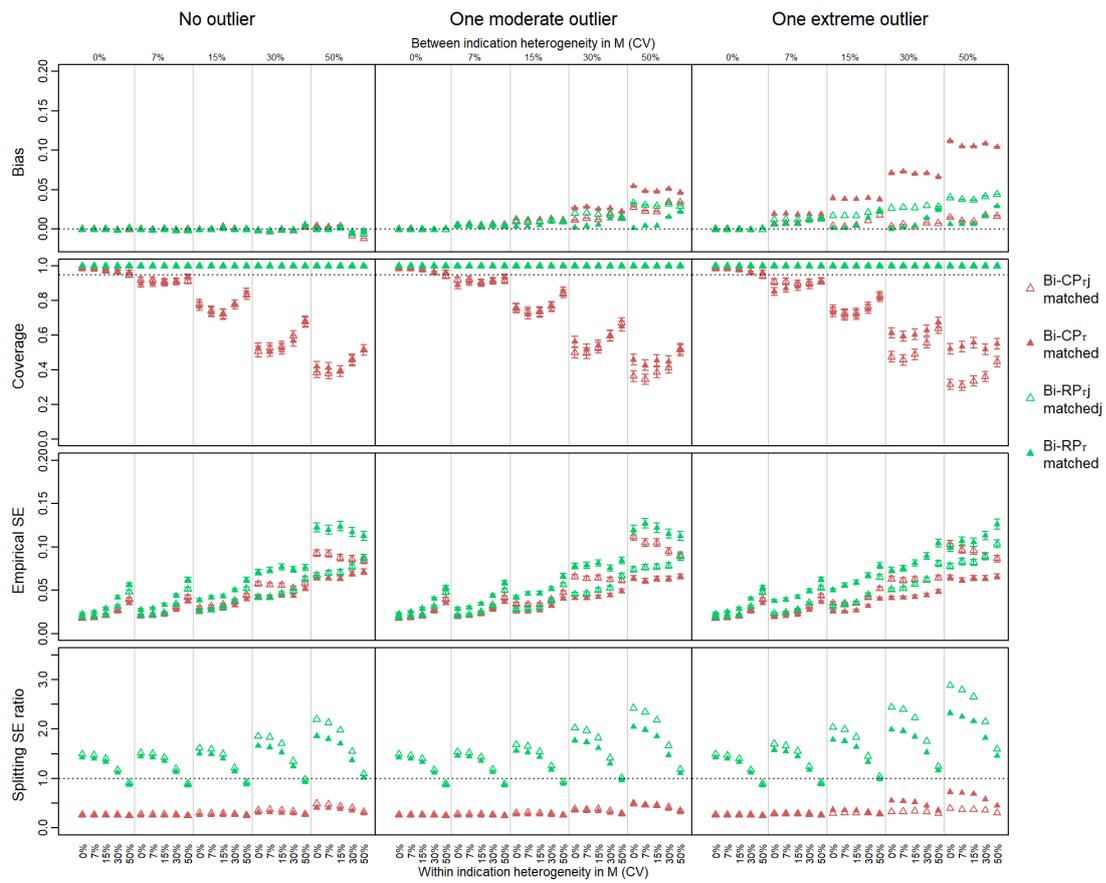

## Small dataset

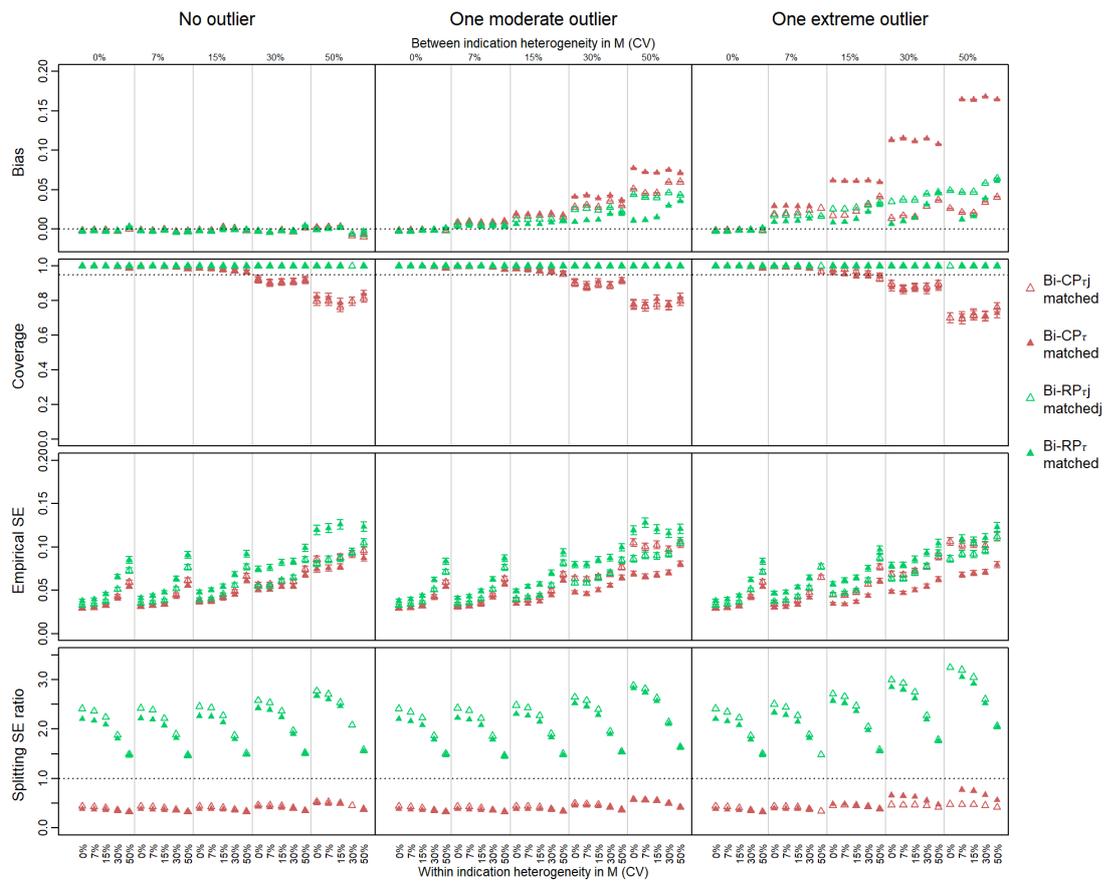

# A4.2 Outlier target indication

Note that the mixture models were not fit to this dataset.

## A4.2.1 With overall survival in target indication

### Univariate non-mixture

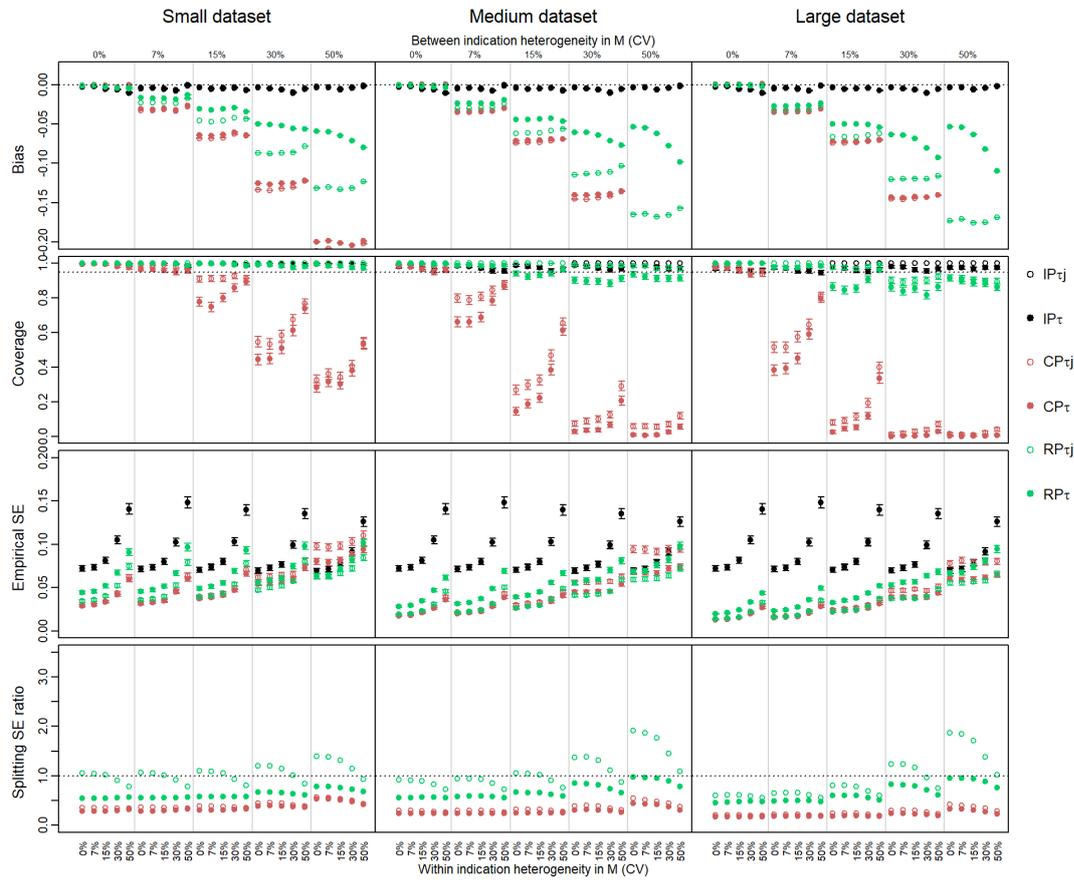

# Surrogate unmatched

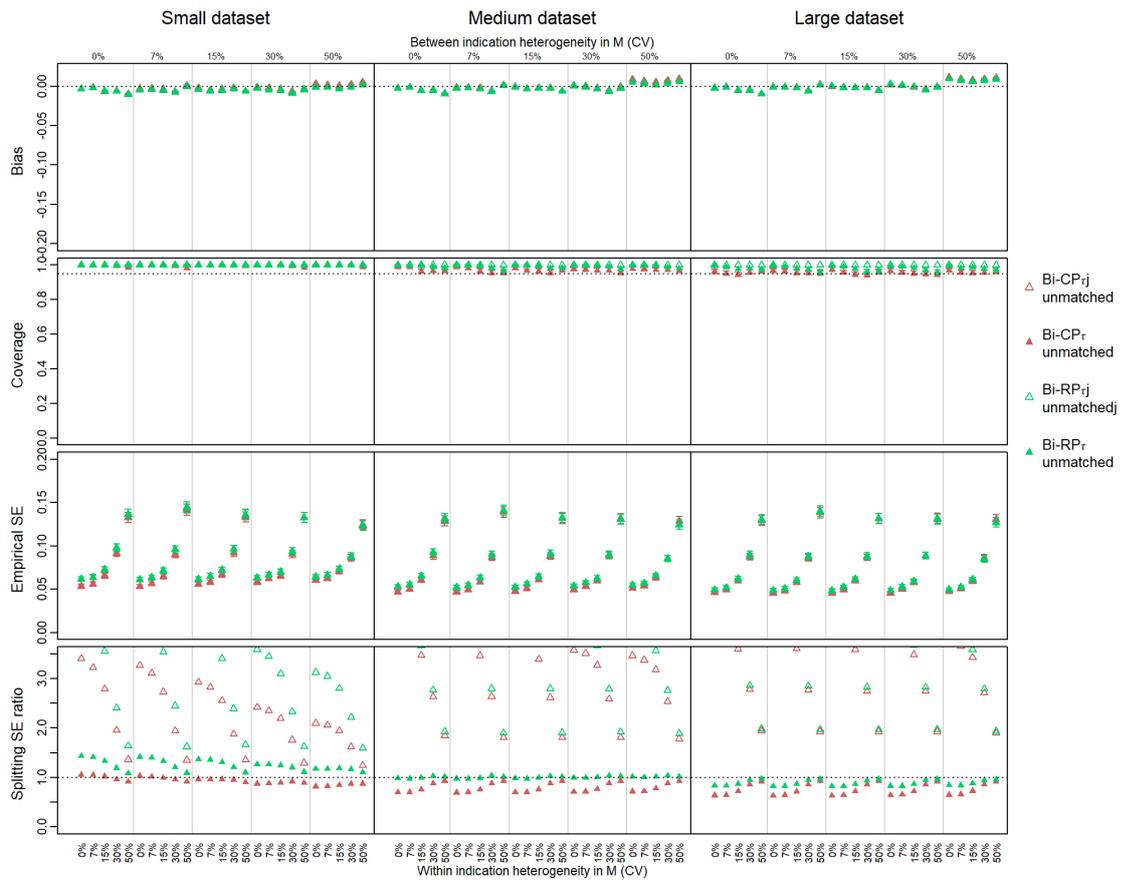

# Surrogate matched

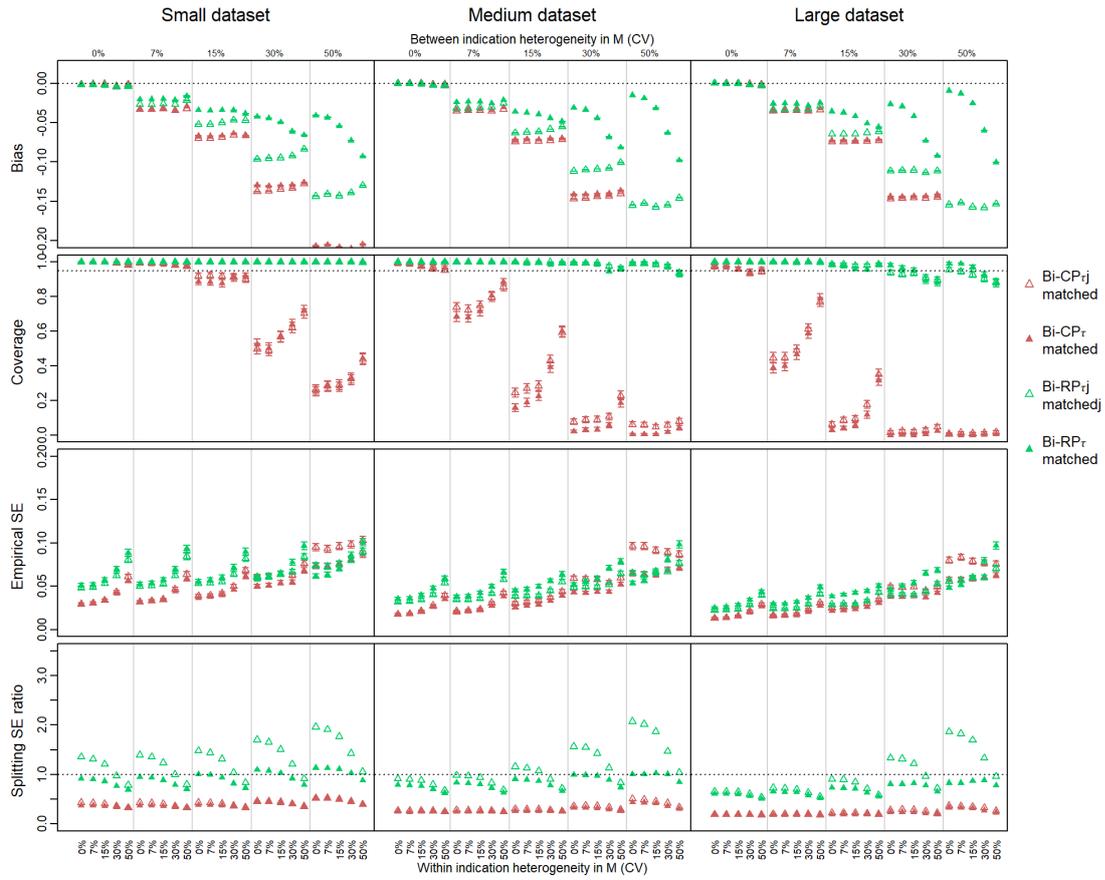

## A4.2.2 No overall survival in target indication

Note that because there is no OS in the target indication it is not possible to estimate the univariate IP model. This causes issues with the interpretation of the splitting SE ratio because this is based on the IP model in the target indication. In the graphs below the splitting SE ratio is calculated comparing the SE of the sharing models to the SE of the IP model if there was OS in the target indication.

### Univariate non-mixture

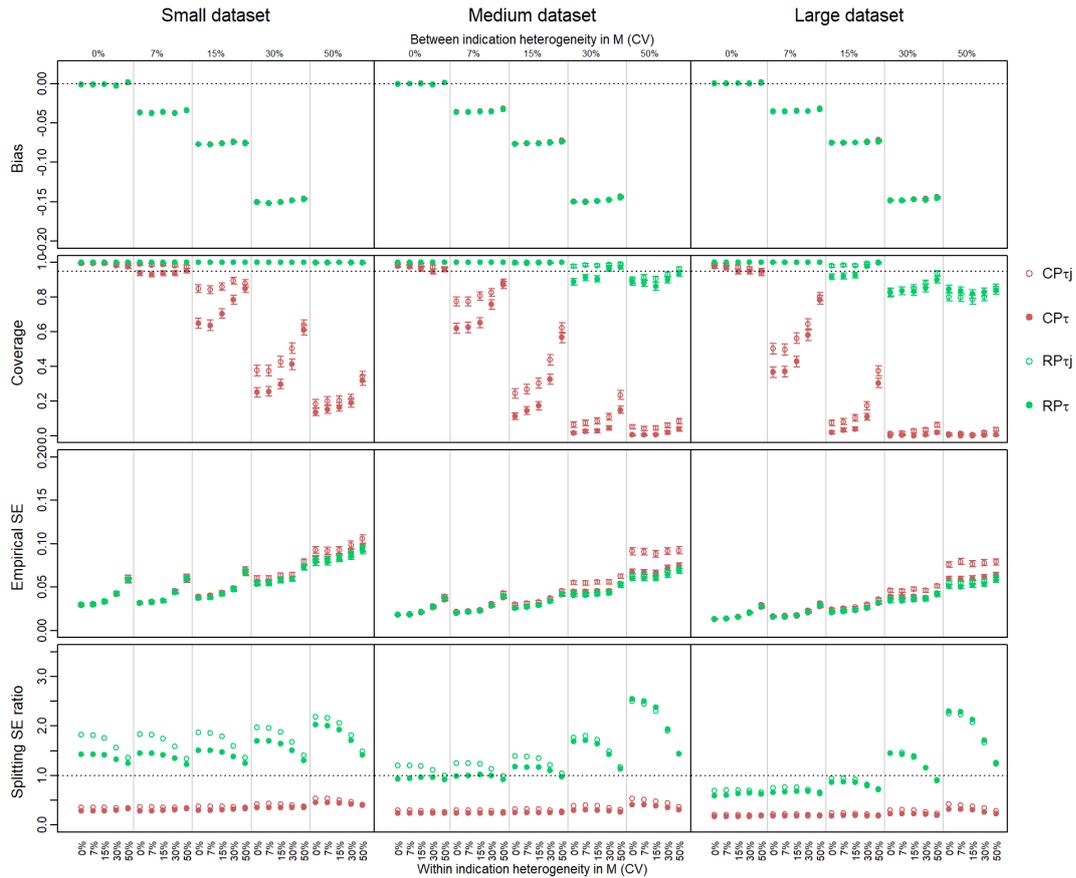

# Surrogate unmatched

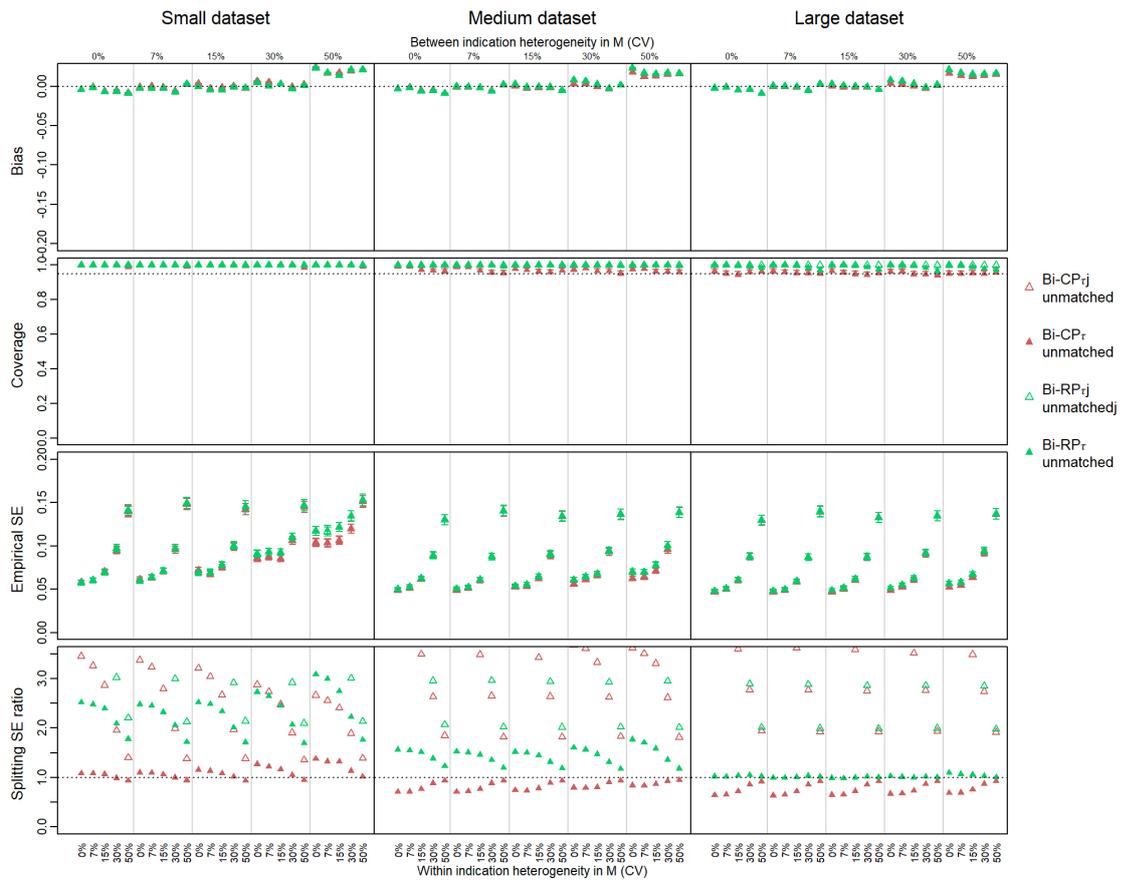

## Surrogate matched

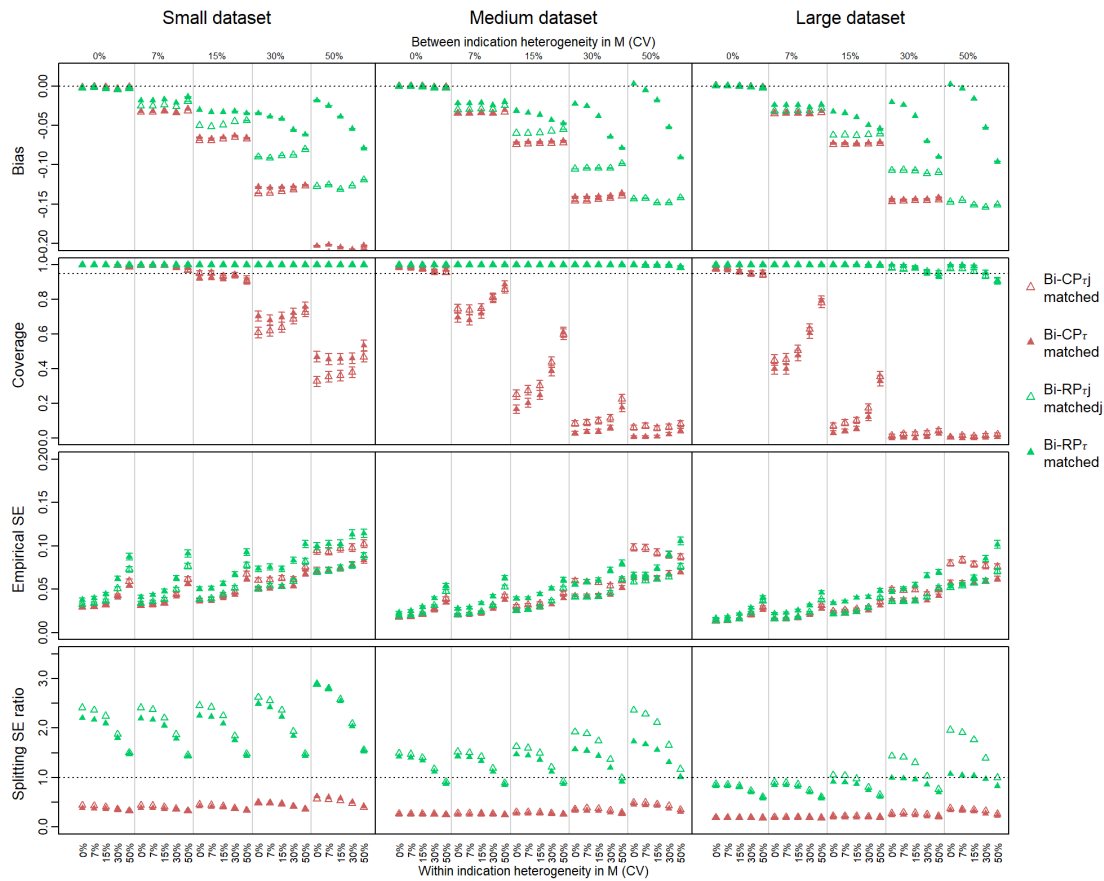

# References

Erdmann, A., J. Beyersmann and K. Rufibach (2025). "Oncology Clinical Trial Design Planning Based on a Multistate Model That Jointly Models Progression-Free and Overall Survival Endpoints." Biometrical Journal **67**(1): e70017.

Jansen, J. P., D. Incerti and T. A. Trikalinos (2023). "Multi-state network meta-analysis of progression and survival data." Statistics in medicine **42**(19): 3371-3391.

Singh, J., S. Anwer, S. Palmer, P. Saramago, A. Thomas, S. Dias, M. Soares and S. Bujkiewicz (2023). "Multi-indication evidence synthesis in oncology health technology assessment." arXiv preprint arXiv:2311.12452.